\documentclass[aps,prb,superscriptaddress,twocolumn,showpacs,amsmath,floatfix,citeautoscript]{revtex4-2}

\usepackage{graphicx}
\usepackage{subfigure}
\usepackage{siunitx}
\usepackage{bm}
\usepackage[colorlinks=true, citecolor=black, urlcolor=black, linkcolor=black]{hyperref}     

\newcommand{\bfr}{ {\bf r}} 
\newcommand{\bfR}{ {\bf R}} 
\newcommand{\bfk}{ {\bf k}} 

\begin{document}

\hyphenpenalty=5000

\tolerance=1000

\title{Density functional theory plus dynamical mean field theory within 
the framework of linear combination of numerical atomic orbitals: Formulation and benchmarks}

\author{Xin Qu}
\affiliation{Rocket Force University of Engineering, Xi'an, Shaanxi 710025, China}
\affiliation{CAS Key Laboratory of Quantum Information, University of Science and Technology of
China, Hefei, Anhui 230026, China}

\author{Peng Xu}
\affiliation{Rocket Force University of Engineering, Xi'an, Shaanxi 710025, China}

\author{Rusong Li}
\affiliation{College of Nuclear Science and Technology, Xi'an Jiaotong University, Xi'an, Shaanxi 710049, China}

\author{Gang Li}
\email{ligang@shanghaitech.edu.cn }
\affiliation{\mbox{School of Physical Science and Technology, ShanghaiTech University, Shanghai 201210, China}}
\affiliation{\mbox{ShanghaiTech Laboratory for Topological Physics, ShanghaiTech University, Shanghai 201210, China}}

\author{Lixin He}
\email{helx@ustc.edu.cn}
\affiliation{CAS Key Laboratory of Quantum Information, University of Science and Technology of
China, Hefei, Anhui 230026, China}

\author{Xinguo Ren}
\email{renxg@iphy.ac.cn}
\affiliation{Institute of Physics, Chinese Academy of Sciences, Beijing 100190, China}
\affiliation{Songshan Lake Materials Laboratory, Dongguan 523808, Guangdong, China}

\date{\today }


\begin{abstract}
The combination of density functional theory with dynamical mean-field theory (DFT+DMFT) has become a powerful first-principles approach to
tackle strongly correlated materials in condensed matter physics. The wide use of this approach relies on robust and easy-to-use implementations,
and its implementation in various numerical frameworks will increase its applicability on the one hand and help crosscheck the validity of the
obtained results on the other. In the work, we develop a formalism within the linear combination of numerical atomic orbital (NAO) basis set framework, which allows
for merging NAO-based DFT codes with DMFT quantum impurity solvers. The formalism is implemented by interfacing two NAO-based DFT codes with
three DMFT impurity solvers, and its validity is testified by benchmark calculations for a wide range of strongly correlated materials, including 
3\textit{d} transition metal compounds, lanthanides, and actinides. Our work not only enables DFT+DMFT calculations using popular and
rapidly developing NAO-based DFT code packages, but also facilitates the combination of more advanced beyond-DFT methodologies available in
this codes with the DMFT machinery.
\end{abstract}

\maketitle
\date{\today}

\section{INTRODUCTION}
\par The understanding of strong correlations among electrons in realistic materials is of great importance for both
fundamental science and technological applications. The strong electron correlations trigger a variety of exotic phenomena, 
which provide a fruitful avenue for designing novel materials~\cite{Kent2018,Paul2019,Adler2019}. 
Kohn-Sham (KS) density functional theory (DFT)~\cite{Hohenberg1964, Kohn1965} under local-density approximation (LDA) 
and generalized gradient approximation (GGA) has achieved remarkable success in describing a very wide range of materials. However, these
approximations are found to be inadequate for correctly describing strongly correlated materials, e.g., 3\textit{d} 
transitional metal compounds, lanthanides, actinides, quantitatively or even qualitatively~\cite{Anisimov1991,Georges1996,Kotliar2006,Held2007}. 
To overcome this limitation, the so-called DFT+DMFT method that merges DFT and many-body technique based
dynamical mean-field theory (DMFT) has been developed \cite{Metzner1989,Ohkawa1991,Georges1992,Jarrell1992,Georges1996,Anisimov1997_DMFT,
Lichtenstein1998,Kotliar2006,Held2007}. Various studies have shown the prominent strength of DFT+DMFT in describing 
electronic structures of strongly correlated materials. DFT+DMFT has thus becoming a promising method of choice in 
studying realistic strongly correlated materials with certain predictive power.~\cite{Savrasov2001,Dai2003,Shim2007-nature,Shim2007-science,Kotliar2006,Held2007}

\par Comparing to DFT+\textit{U} \cite{Anisimov1991,Anisimov1993,Anisimov1997_LDAU}, a method that shares similar spirit
as DFT+DMFT and is available in nearly  all popular DFT code packages, DFT+DMFT is not only computationally more expensive, but also 
relatively more difficult to access for normal users. 
Implementing DFT+DMFT with well-defined local orbital basis and in a way that doesn't requires much 
expertise from users is, thus, of practical importance for promoting the methodology in studying realistic 
correlated materials. One of the key issues in the implementation of DFT+DMFT scheme is the 
definition of local correlated orbitals, in terms of which the DMFT impurity problem is defined and solved. Many-body corrections  
to the KS Hamiltonian exist only in this subspace. As a result, the choice of local correlated orbitals has a noticeable influence on
the obtained DFT+DMFT results at the quantitative level.

\par In early days, DFT+DMFT was implemented within the linear muffin-tin orbital (LMTO) basis set framework
\cite{Anisimov1997_DMFT, Lichtenstein1998, Savrasov2001,Pourovskii2007,Granas2012}, where the local LMTOs were chosen as the basis orbitals hosting
strongly correlated \textit{d} or \textit{f} electrons. Later, different Wannier-type orbitals, such as projective Wannier
functions \cite{Anisimov2005} and maximally localized Wannier functions, were used to define the local orbitals \cite{Pavarini2004,Lechermann2006}.
As for plane-wave based DFT, Amadon  \textit{et al.} \cite{Amadon2008-dmft,Amadon2012} used the all-electron atomic partial waves
within the projector augmented wave (PAW) framework or pseudoatomic wave functions as the local correlated orbitals.
Within the linearized augmented plane-wave (LAPW) method,  Aichhorn \textit{et al.} \cite{Aichhorn2009}, and Haule \textit{et al.} and 
Kim \cite{Haule2010} independently developed projector schemes that convert the KS orbital space to local correlated subspace.
We note that, in recent years, the underlying embedding idea of DMFT has been used in broader sense whereby the embedded cluster is solved
in a \textit{ab initio} way using beyond-DFT approaches such as $GW$ or coupled-cluster method, whereas the environment is treated
using less advanced approaches~\cite{Zgid2011,Chibani2016,Zhu2021}.

\par In recent years, numerical atomic orbitals (NAOs) have been employed as the basis set choice in several first-principles software packages \cite{Soler2002a,Ozaki/etal:2008,Delley2000,Blum2009,Li2016}.
Unlike most other basis sets, the linear combination of NAOs represents a versatile framework that 
can be used in both all-electron and pseudopotential-based DFT calculations. Past experience indicates
that NAOs are an efficient basis set choice not only for conventional ground-state calculations, but also for functionalities that go beyond
conventional DFT calculations whereby explicit two-electron Coulomb integrals and/or 
unoccupied KS states are needed \cite{Ren2012,Levchenko2015,Lin2021,Ren2021}. In this regard, it is highly desirable to develop computational schemes 
within the NAO basis set framework that enables a convenient merging of first-principles and model-Hamiltonian based approaches. 
In case of DFT+$U$, several NAO-based implementations have been reported previously \cite{Han2006,Kick2019,DFTU-arxiv}.
However, in case of DFT+DMFT, only one recent work within the pseudoatomic orbital basis sets (i.e., NAOs tailored for
pseudopotential-based calculations) was reported by Sim and Han \cite{Sim2019}. In their work, these authors proposed to use
the so-called natural atomic orbitals -- the eigen-orbitals for the local density matrix -- to define the local correlated orbitals.
In the present work, we develop a projection scheme that allows to conveniently interface the NAO-based DFT codes and the DMFT impurity solvers
and thus enables NAO-based DFT+DMFT calculations. We test the efficacy of such a scheme for the all-electron FHI-aims code \cite{Blum2009} and
the norm-conserving pseudopotential based ABACUS code \cite{Li2016}, interfaced with three different DMFT impurity solver packages. 
Consistent results are obtained for transition metal compounds, including both correlated metals (SrVO$_3$) and Mott insulators (NiO and MnO), 
rare-earth systems (Ce metal and Ce$_2$O$_3$), and actinides (Pu$_2$O$_3$ and PuO$_2$). Our work thus extends the reach of the NAO-based
numerical framework to treat strongly correlated materials.

\par This paper is organized as follows. Sec. \ref{Sec2} focuses on the DFT+DMFT formalism. After introducing the general 
DFT+DMFT formulation, we present our definition of local correlated orbitals within the NAO framework.
This is followed by a discussion of the self-consistency scheme of our DFT+DMFT implementation. In Sec. \ref{Sec3}, we benchmark 
the validity of our DFT+DMFT formalism and implementation over a wide range of strongly correlated materials, i.e., 3\textit{d}-systems
SrVO$_3$, MnO and NiO, 4\textit{f}-systems Ce metal and Ce$_2$O$_3$, and 5\textit{f}-systems Pu$_2$O$_3$ and PuO$_2$. Sec.~\ref{Sec4} concludes
this paper.

\section{THEORETICAL FORMULATION}\label{Sec2}
\subsection{General DFT+DMFT formalism}
\par Combining DFT with DMFT is not straightforward conceptually as the two theories are formulated in different forms and in different Hilbert spaces. 
In KS-DFT, one needs to self-consistently solve an effective single-particle problem, whereby (for periodic systems) a KS Hamiltonian is constructed and solved
separately at each individual $\bfk$ point in the Brillouin zone (BZ).
On the other hand, DMFT is a real-space approach developed to solve the lattice Hamiltonian models in which the on-site two-electron interactions are explicitly included.
Furthermore, as a first-principles approach, DFT accounts for all the chemical details in the systems and include all (at least valence) electrons
in the calculations. In contrast, only the strongly correlated electrons sitting  energetically in the vicinity of the Fermi level, relevant for the low-energy physics, are  included in the model Hamiltonian and treated explicitly in DMFT.
As such, key quantities like charge density, density matrix, and Green functions all have different representations in DFT and DMFT, 
impeding a straightforward combination of the two approaches. One solution to address this challenge, as detailed below, is to define a suitable projector
to convert the key quantities back and forth between the two representations. As a side note, in the following we only focus on the general formalism of 
DFT+DMFT, without going into a detailed account of DMFT itself in this paper. Interested readers are referred to 
Refs. \cite{Metzner1989,Georges1992,Georges1996,Kotliar2006,Held2007,Vollhardt2010}  for a detailed discussion of this theory. 

\par Briefly, in the DFT+DMFT formalism, the Green functions defined in the correlated subspace are the central objects gluing the two theories. 
There are two kinds of interacting Green functions defined in the corrected subspace:
One is the impurity Green function $G^{imp}$ determined by solving the impurity problem in Anderson impurity model, describing a single site with on-site Coulomb interactions and coupled to a mean-field electron bath. The another one, denoted as $G^{loc}$, is the on-site term of the
lattice Green function, which is essentially the the Green function of the correlated site described by the Hubbard model
and is obtained by projecting the Green function in the 
full Hilbert space into the correlated subspace. The fundamental requirement in DMFT is that the ``on-site" Green function of 
the lattice problem and that of the Anderson impurity model are equal,
\begin{equation}\label{Eq1}
	G^{loc} = G^{imp}\, ,
\end{equation}
which is achieved via the DMFT self-consistency cycle.

\par In the DFT+DMFT scheme, the single-particle effective Hamiltonian in the full Hilbert space can be expressed as 
\begin{equation}\label{Eq2}
  \hat{H} = \hat{H}_\mathrm{KS} + \hat{\Sigma} - \hat{H}_{\mathrm{dc}} \, ,
\end{equation}
where $\hat{H}_\mathrm{KS}$ is the non-interacting KS Hamiltonian, $\hat{\Sigma}$ the self-energy which
encodes all the many-body complexities arising from strong electron correlations and are nonzero only for correlated orbitals. 
Finally, $\hat{H}_{\mathrm{dc}}$ is the double-counting term, which is introduced to discount the interactions 
among correlated electrons that have been included in a static mean-field level in $\hat{H}_\mathrm{KS}$.

\par Starting from the Hamiltonian Eq.~(\ref{Eq2}), the key issue in the combination of DFT and DMFT is the downfolding (projecting) of
the physical quantities from the full KS space to the local correlated space, and the upfolding (embedding) of these quantities
from the local space to the full space. Mathematically, the local interacting Green function $G^{loc}$ can be obtained through a 
projection procedure. In the literature, two different projection procedures have been used. The first one can be seen
as a ``Hamiltonian projection", whereby a tight-binding Hamiltonian is first obtained from the KS Hamiltonian by projecting 
the latter into the correlated subspace,
\begin{equation}\label{Eq3}
  H_{m m^\prime}^{TB}(\bfk) = \langle \phi_{m\bfk} | \hat{H}_\mathrm{KS} | \phi_{m^\prime\bfk} \rangle \, .
\end{equation}
Here $\phi_{m\bfk}$ is the Bloch-summed local orbitals with $m$ denoting an orbital index in the correlated subspace.
In the second step, the local interacting Green function can be directly calculated from this tight-binding Hamiltonian as
\begin{equation}\label{Eq4}
  \begin{aligned}
    G_{mm^\prime}^{loc}(i\omega_n) =& \sum_{\bfk} \frac{1}{N_\bfk} \left[ (i\omega_n + \mu)I - H^{TB}(\bfk) \right. \\
    & \left. - \Sigma (i\omega_n) + H_{\mathrm{dc}} \right]_{mm'}^{-1} \, ,
  \end{aligned} 
\end{equation}
where $N_\bfk$ is the is the number of $\bfk$ points in the first Brillouin zone, equivalent to the number of unit cells in the
Born-von K\'{a}rm\'{a}n supercell. In Eq.~(\ref{Eq4}), $\mu$ is the chemical potential, $I$ an identity matrix, and $[\cdots]^{-1}$ denotes the 
matrix inversion which is taken within the correlated subspace.
Here and below, we employ the Matsubara Green function formalism where $\omega_n=(2n+1)\pi/\beta$ are the discrete
frequency points along the imaginary frequency axis, with $\beta=1/k_B T$ being the inverse temperature. 
The spectroscopy results on the real frequency axis are obtained via analytical continuations. 
The above procedure to compute the local interacting Green function is the so-called tight-binding Hamiltonian method. 

\par In the second scheme, one first constructs the interacting Green function in the full Hilbert space as
\begin{equation}\label{Eq5}
  G_{ij}(\bfk,i\omega_n) = \left[ (i\omega_n + \mu) I - H(\bfk)  \right]_{ij}^{-1}\, ,
\end{equation}
where the Hamiltonian matrix $H_{ij}(\bfk) = \langle \chi_{i,\bfk} \mid \hat{H} \mid \chi_{j,\bfk}\rangle$ is the representation 
of the interacting Hamiltonian operator introduced in Eq.~(\ref{Eq2}) within an arbitrary orthonormal basis set $\mid \chi_{j,\bfk}\rangle$ 
expanding the full Hilbert space. While the matrix elements of the KS Hamiltonian $\hat{H}_\mathrm{KS}$ within such a basis set 
are given straightforwardly, those of $\hat{\Sigma}$ and  $\hat{H}_{\mathrm{dc}}$ will be discussed later in Sec.~\ref{Sec2c}.
Once the full-space lattice Green function is obtained, the local interacting Green function can be obtained through a projection procedure
\begin{equation}\label{Eq6}
  G_{mm^\prime}^{loc}(i\omega_n) = \sum_{\bfk} \frac{1}{N_\bfk} \sum_{ij} P_{ij}(\bfk, m m^\prime) G_{ji}(\bfk,i\omega_n) ~, 
\end{equation}
where the projector above is given by
\begin{equation}\label{Eq7}
  P_{ij}(\bfk, m m^\prime) = \langle \phi_m | \chi_{i\bfk}\rangle 
  \langle \chi_{j\bfk} | \phi_{m^\prime} \rangle \, .
\end{equation}
The construction of the local Green function via Eqs.~(\ref{Eq5})-(\ref{Eq7}) is known as the projector method.

\par The two methods discussed above to construct the local Green function 
differ in that the tight-binding Hamiltonian method projects the Hamiltonian matrix while the projector 
method projects the Green function. In the projector method, the matrix inversion 
is carried out in the full Hilbert space, as indicated in Eq.~(\ref{Eq5}), and thus the interactions between the 
correlated electrons and the rest electrons are retained to some extent. On the other hand, within the tight-binding method, 
if one wants to describe the interaction between correlated electrons and the rest ones, e.g., studying the charge transfer process
between the correlated orbitals and the bath, one has to enlarge the Hilbert space of the tight-binding
Hamiltonian to encompass the extra itinerant electrons, and it will inevitably increases the computational complexity. 
In this work, we choose the projector method in our DFT+DMFT implementation.

\par The other key quantity in DFT+DMFT calculations is the interacting impurity Green function
$G^{imp}$ in Eq.~(\ref{Eq1}), corresponding to the local propagator of the effective single-impurity Anderson model,
which describes a single site coupled to a bath that mimics the lattice environment at a mean-field level.
Formally $G^{imp}$ satisfies the following relationship
\begin{equation}\label{Eq8}
 \begin{aligned}
  G^{imp}_{mm^\prime} \left(i\omega_n\right) =& [ (i\omega_n + \mu)I - {\cal E}^{imp} - \Delta\left(i\omega_n\right) \\ 
  & - \Sigma^{imp}\left(i\omega_n\right)]_{mm^\prime}^{-1}\, ,
  \end{aligned}
\end{equation}
where ${\cal E}^{imp}$ is the impurity energy level,  $\Delta\left(i\omega_n\right)$ the so-called hybridization function 
characterizing the influence of the environment on the embedded impurity, and $\Sigma^{imp}\left(i\omega_n\right)$  the impurity
self-energy. It should be noted that ${\cal E}^{imp}$ and $\Delta\left(i\omega_n\right)$ are matrices when there are multiple orbitals 
in the correlated subspace, which is typical in DFT+DMFT calculations. In this context, it is also customary to define a so-called 
Weiss Green function
\begin{equation}\label{Weiss-filed}
  \begin{aligned}
      \mathcal{G}_{mm^\prime}^{-1}(i\omega_n) =& G_{mm^{\prime}}^{-1}(i\omega_n) + \Sigma_{mm^\prime} \\
      =&  [(i\omega_n + \mu)I - {\cal E}^{imp} - \Delta\left(i\omega_n\right) ]_{mm^\prime} ~,
  \end{aligned} 
\end{equation}
which acts as the dynamical (energy-dependent) mean field that the impurity electrons experience, and encodes essentially 
the same information as the hybridization function $\Delta\left(i\omega_n\right)$. When the self-consistency in the DMFT loop is reached,
the local Green function $G_{mm^{\prime}}^{loc}(i\omega_n)$ and the local self-energy $\Sigma_{mm^\prime}(i\omega_n)$ in Eq.~(\ref{Eq4}) 
and (\ref{Eq6}) are equal to the impurity Green function $G^{imp}_{mm^{\prime}}(i\omega_n)$ and the impurity self-energy 
$\Sigma^{imp}_{mm^\prime}(i\omega_n)$, respectively.

\par The Weiss Green function together with the local Coulomb interactions defines the Anderson impurity model, which can be expressed in terms of an action
\cite{Georges1996}
 \begin{equation}\label{Eq9}
  \begin{aligned}
     S=&\int_{0}^{\beta} d\tau \sum_{m m^\prime} c_{m}^{\dagger}(\tau) \mathcal{G}_{mm^\prime}^{-1}\left(\tau\right)c_{m^\prime}\left(\tau\right)\\
       & -\sum_{l m n o} U_{l m n o} \int_{0}^{\beta} d \tau c_{l}^{\dagger}(\tau) c_{n}(\tau) c_{m}^{\dagger}(\tau) c_{o}(\tau) ~,
   \end{aligned} 
 \end{equation}
where  $c_{l}^{\dagger}(\tau)$, $c_{n}(\tau)$, etc., should be understood as the Grassmann variables, and $U_{l m n o}$ is the on-site Coulomb interaction
expressed within a set of local orbitals (labelled by $l,m,n,o$) spanning the correlated subspace. The action $S$ is 
essentially the integration of the Lagrangian over the imaginary time. For given $S$, the impurity Green function can be calculated as
 \begin{equation}\label{Eq10}
  G^{imp}_{m m^\prime} = -\frac{1}{\mathcal{Z}} \int \mathcal{D} \prod_{i} \left[ c^{\dagger}, c \right] c_m c_{m^\prime}^\dagger e^{-S} ~, 
 \end{equation}
 where $i$ runs over all $m$ indices, and $\mathcal{Z}$ is the partition function 
 \begin{equation}\label{Eq11}
   \mathcal{Z} =  \int \mathcal{D} \prod_{i} \left[c^{\dagger}, c\right] e^{-S} \, .
 \end{equation}
The interacting impurity Green's function defined via Eqs.~(\ref{Eq9}-\ref{Eq11})
can then be obtained through a variety of numerical approaches -- usually termed as impurity solvers. 
Up to date, several types of impurity solvers have been developed,
including the quantum Monte Carlo (QMC) \cite{Jarrell1992,Hirsch1986,Rozenberg1992,Georges1992-qmc}, 
non-crossing approximation (NCA) \cite{Pruschke1993a, Pruschke1993b, Jarrell1994}, one-crossing approximation (OCA) \cite{Haule2001,Haule2005}, 
exact diagonalization (ED) \cite{Si1994, Caffarel1994}, numerical renormalization group (NRG) \cite{Sakai1994, Bulla1998}, etc. 
Among all these impurity solvers, continuous-time quantum Monte Carlo (CTQMC) \cite{Werner2006, Gull2007, Haule2007, Gull2011a}
provides access to both high and low energy scales and is effective for a wide class of realistic material calculations. 
Nowadays, the CTQMC,   especially the hybridization expansion based CTQMC (CT-HYB), is the most popular impurity solver employed in DFT+DMFT
calculations.

\par Once the interacting impurity Green function is determined, the impurity self-energy $\Sigma^{imp}(i\omega_n)$ can be obtained
via the Dyson equation $\Sigma^{imp}(i\omega_n)= \left[\mathcal{G}(i\omega_n)\right]^{-1} -\left[G^{imp}(i\omega_n)\right]^{-1}$, 
which is usually done within the impurity solvers. This impurity self-energy will be taken as the updated local self-energy and 
fed into Eq.~(\ref{Eq3}) and (\ref{Eq4}) or (\ref{Eq2}) and (\ref{Eq5}), from which a new local Green function and consequently 
a new Weiss Green's function can be obtained. This is where a new iteration starts. This self-consistency loop keeps going until 
the self-energy reaches convergence or the local and impurity interacting Green function satisfies the self-consistency condition, 
i.e., Eq.~(\ref{Eq1}).

\subsection{Construction of the projector}
\par Within the NAO basis set framework, it's natural to take the $d$ or $f$-type NAOs that contribute 
most to the electronic states around the Fermi level as the the local correlated orbitals. These NAOs by construction are localized and atom-centered, 
and thus satisfy the usual requirement of correlated orbitals. In early DFT+DMFT implementations, analogous atomic-like 
orbitals like LMTOs were used. LMTOs are minimal basis sets in the sense that for each angular momentum channel there is only
one radial basis function. In contrast with LMTOs, the NAO basis sets are of multi-$\zeta$ character meaning that there are more than one radial functions
per angular momentum, thus offering a more accurate description of the electronic
structure. In the past, the DFT+\textit{U} method has been successfully implemented within NAO-based DFT codes~\cite{Han2006, Kick2019, DFTU-arxiv},
whereby it turns out to be a good practice to choose the most localized $d$ or $f$ basis functions as the correlated orbitals to apply the Hubbard $U$ correction.
Thus the most localized $d$ or $f$ orbitals seems to span the correlated subspace rather well.  In practice, since NAOs
centering on neighboring atoms are non-orthogonal to each other, certain orthorgonalization procedure is needed to generate
an orthonormal local basis set, which is convenient for DMFT calculations. 

\par Below we shall discuss our procedure to construct the projector and local correlated orbitals to facilitate DFT+DMFT calculations
within the NAO basis set framework. In analogy to the transformation between Bloch orbitals and Wannier orbitals, 
we define the following Bloch-summed atomic orbitals as
\begin{equation} \label{Eq13}
  \Phi_{I,m}^{\bf{k}}(\bfr) = \frac{1}{\sqrt{N_\bfk}} \sum_{\mathbf{R}} e^{i\mathbf{k}\cdot\mathbf{R}} \phi_{I,m}\left(\bfr -
  {\bm \tau}_{I}-\bf{R}\right)  \, ,
\end{equation}
where $\phi_{I,m}\left(\bfr - {\bm \tau}_{I}-\bf{R}\right)$ is a NAO located at atomic site $I$ in cell $\bfR$.
Here the magnetic quantum number $m$ labels the different orbitals in the correlated angular moment channel, such as
the five $d$ orbitals or seven $f$ orbitals. 

\par Since NAOs on neighboring atomic sites are non-orthogonal to each other, it is obvious that the $\Phi_{I,m}^{\bfk}(\bfr)$'s 
defined in Eq.~(\ref{Eq13}) with different $m$ are also non-orthogonal. The key next step is to employ the L$\ddot{\text{o}}$wdin orthonormalization procedure to $\Phi_{I,m}^{\mathbf{k}}$, i.e., 
\begin{equation} \label{Eq14}
  \mid \tilde{\Phi}_{I,m}^{\mathbf{k}} \rangle = \sum_{I^\prime m^\prime} O^{-\frac{1}{2}}_{Im, I^\prime m^\prime} 
  \left( \mathbf{k} \right) \mid \Phi_{I^\prime,m^\prime}^{\mathbf{k}} \rangle \, ,
\end{equation}
where 
\begin{equation} \label{Eq15}
  O_{Im, I^\prime m^\prime} \left( \mathbf{k} \right) = \langle \Phi_{I,m}^{\mathbf{k}} \mid 
  \Phi_{I^\prime ,m^\prime}^{\mathbf{k}} \rangle
\end{equation}
is the overlap matrix. The newly obtained $\tilde{\Phi}_{I,m}^{\mathbf{k}}$ orbitals are then orthonormal by construction. 
Afterwards, a Fourier transform is applied to get the correlated orbital in real space, i.e.,
\begin{equation} \label{Eq16}
  \mid W_{I,m}^{\bfR} \rangle = \frac{1}{\sqrt{N_\bfk}}\sum_{\mathbf{k}} e^{-i\bfk \cdot \mathbf{R}} \mid \tilde{\Phi}_{I,m}^{\mathbf{k}} \rangle ~.
\end{equation}
The orthonormality of $W_{I,m}^{\bfR}$ is also guaranteed by construction.

\par We choose the KS states $|\Psi_{i\bfk}\rangle$  
as the basis sets $|\chi_{i\bfk}\rangle$ (cf. Eq.~(\ref{Eq7})) to span the full Hilbert space, and then Eq.~(\ref{Eq6}) becomes
\begin{equation}\label{Eq17}
  \begin{aligned}
  G_{I, mm^\prime}^{loc} =& \sum_{\mathbf{k}} \frac{1}{N_\bfk} \sum_{ij} P^I_{ij}( \mathbf{k}, m m^\prime) \\
  & \left[ \frac{1}{i\omega_n + \mu - \epsilon({\mathbf{k}}) - \bar{\Sigma}( \mathbf{k}, i\omega_n) } \right]_{ji} ,
  \end{aligned} 
\end{equation}
where $ \epsilon_{ji}({\mathbf{k}}) = \langle \Psi_{j \mathbf{k}}|\hat{H}_\textrm{KS}|\Psi_{i \mathbf{k}} \rangle =  \epsilon_{i\bfk} \delta_{ij} $ and  $\bar{\Sigma}_{ji}(\mathbf{k},i\omega_n) = \langle \Psi_{j \mathbf{k}}| \hat{\Sigma}(i\omega_n) - \hat{H}_{\mathrm{dc}}|\Psi_{i \mathbf{k}} \rangle$, with $\epsilon_{i\bfk}$ being the KS eigenvalues. 
The projector, Eq.~(\ref{Eq7}), then becomes
\begin{equation}\label{Eq18}
  P_{ij}^{I}\left(\bfk,m m^\prime\right) =  \langle \Psi_{i\mathbf{k}} \mid W_{I,m}^{0} \rangle 
  \langle  W_{I,m^\prime}^{0}\mid \Psi_{j\mathbf{k}}\rangle \, ,
\end{equation}
where the superscript $0$ denotes the central unit cell. Since in DMFT calculations, only the ``on-site" Green function, where
$m$ and $m'$ orbitals are located in the same unit cell is needed, the projector is designed to project the full Green function
into the central unit cell, without losing generality.
Formally, the projector for a given correlated atom $I$ and a wave vector $\bfk$ is a fourth-order tensor, but since
it is separable and symmetric, only a second-order tensor, i.e., the overlap matrix $\langle \Psi_{i\mathbf{k}} \mid W_{I,m}^{0} \rangle$ 
needs to be computed and stored in practical implementations.

\par The whole DFT+DMFT scheme requires the orthonormality of local orbitals representing the correlated subspace. 
In the language of the projector, it requires the projector to satisfy the following orthonormal condition
\begin{equation}\label{Eq19}
  \begin{aligned}
  \sum_{i} P_{ii}^{I}\left(\bfk,m m^\prime\right) =& \sum_{i} \langle W_{I,m^\prime}^{0}\mid \Psi_{i\mathbf{k}}\rangle 
  \langle\Psi_{i\mathbf{k}} \mid W_{I,m}^{0} \rangle \\
  =& \delta_{m m^\prime} .
  \end{aligned}
\end{equation}
In principle, this condition is automatically satisfied if the summation over $i$ goes over all the KS bands.
In practical DFT+DMFT implementations, however, one truncates the full KS states into a small subset around 
the Fermi level, which means $i$ just runs over bands that are located in a chosen energy window around the 
Fermi level (in the following, these subsets of bands are denoted as $\mathcal{C}$). 
This truncation destroys the completeness of $|\Psi_{ik}\rangle$ and thus the orthonormality of the projector. 
To deal with this issue, one can introduce an extra transformation
\begin{equation}\label{Eq20}
 \begin{aligned}
  \tilde{P}_{ij}^{I}\left(\mathbf{k},m m^\prime\right) = & \sum_{m^{\prime \prime \prime}} \tilde{O}_{m^\prime m^{\prime \prime \prime}}^{-\frac{1}{2}}(\mathbf{k})\langle  W_{I,m^{\prime \prime \prime}}^{0}\mid \Psi_{j\mathbf{k}}\rangle \\
  &  \sum_{m^{\prime \prime}}  \langle \Psi_{i\mathbf{k}} \mid W_{I,m^{\prime \prime}}^{0}\rangle \tilde{O}_{m^{\prime \prime} m}^{-\frac{1}{2}}(\mathbf{k})
 \end{aligned}
\end{equation}
to orthonormalize the projector. The transformation matrix in the above equation is given by
\begin{align}\label{Eq21}
  \tilde{O}_{m m^\prime}(\mathbf{k}) & = \sum_{i \in \mathcal{C}} P_{ii}^{I}\left(\mathbf{k},m m^\prime\right) \nonumber \\
  & = \sum_{i \in \mathcal{C}} \langle  W_{I,m}^{0}| \Psi_{i\mathbf{k}}\rangle\langle \Psi_{i\mathbf{k}} | W_{I,m'}^{0} \rangle \, , 
\end{align}
which is nothing but the overlap between the projections of the orthonormalized local orbitals $|W_{I,m}^{0}\rangle$'s 
within the subspace $\mathcal{C}$.
Mathematically, the local correlated orbitals we use above to construct the projector can be explicitly expressed as 
\begin{equation}\label{Eq22}
  \begin{aligned}
  \mid \tilde{W}_{I,m}^{\bfR} \rangle =& \frac{1}{\sqrt{N_{\mathbf{k}}}}\sum_{\mathbf{k}} e^{-i\mathbf{k}{\bfR}} \sum_{m^\prime} \tilde{O}_{m m^\prime}^{-\frac{1}{2}}(\mathbf{k}) \\
  &\sum_{i \in \mathcal{C}} \langle \Psi_{i\mathbf{k}} | \tilde{\Phi}_{I,m^\prime}^{\mathbf{k}} \rangle | \Psi_{i\mathbf{k}} \rangle .
  \end{aligned}
\end{equation}
In this form, our scheme is similar in spirit to the projective Wannier-orbital scheme proposed by Anisimov \textit{et al.}~\cite{Anisimov2005}
in the context of LDA+DMFT calculations. In our case, the most localized $d$ or $f$ NAO plays the role of the LMTO in the work
of Ref.~\cite{Anisimov2005}



\par There are a few advantages of using NAOs to define the local correlated space. Firstly, this choice 
is physically intuitive and technologically straightforward within the NAO basis set framework. 
We do not need to spend extra efforts to generate a set of local orbitals and make sure they are physically 
reasonable atomic-like orbitals. Secondly, from both the theoretical and technical perspectives, the choice 
of NAOs to define the local correlated space and the resulting projection scheme are suitable for all NAO-based packages. 
Especially, the key quantities that are required in this formalism, e.g., the KS wave functions and the overlap matrix 
of the basis functions, exist naturally in NAO-based DFT code packages, and hence no additional 
efforts are required to calculate these quantities. Thirdly, due to its high flexibility, our DFT+DMFT infrastructure 
can be interfaced with a new NAO-based DFT code without much effort. We hope it can serve as a platform to enable NAO-based 
DFT codes to do DFT+DMFT researches on strongly correlated materials. In this work, we implement the DFT+DMFT interface and 
test it with two NAO-based DFT codes using different techniques, i.e., the pseudopotential-based
ABACUS code~\cite{Li2016} and full-potential all-electron FHI-aims code~\cite{Blum2009}. 

\subsection{DFT+DMFT self-consistency scheme}\label{Sec2c}
\par
\begin{figure*}[htbp]
  \centering
  \includegraphics[width=0.8\linewidth]{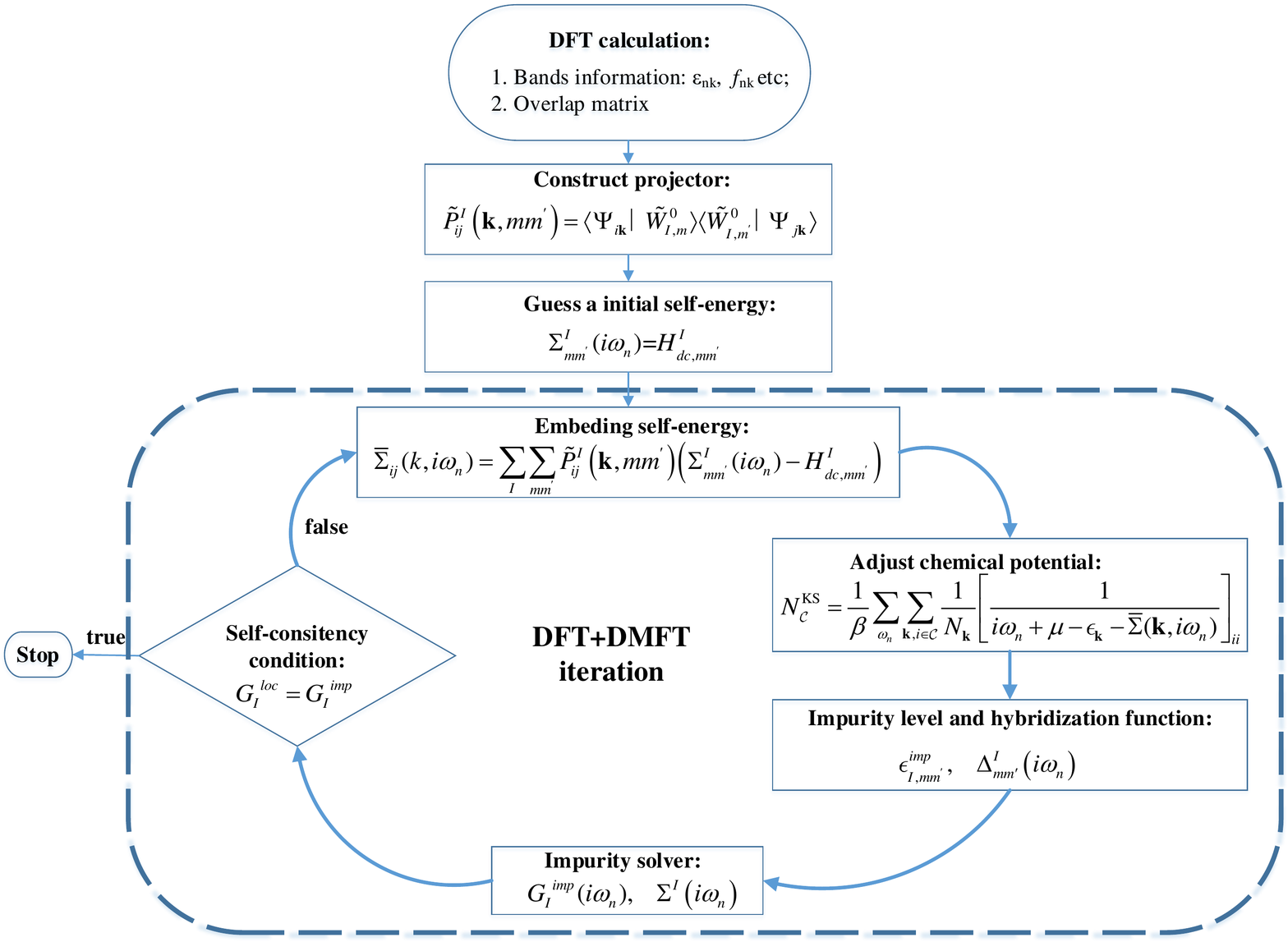}
  \caption{Flow diagram of the major steps in our DFT+DMFT implementation.}
  \label{flow-diagram}
\end{figure*}

 In this section, we will explain step by step our DFT+DMFT calculation procedure, according to the workflow outlined in the flow diagram
depicted in Fig.~\ref{flow-diagram}.

\par \textit{Step} 1.  The DFT+DMFT calculation starts from well-converged DFT band structures. To get high-quality KS orbitals $|\Psi_{i\bf{k}}\rangle$, a dense $\bfk$-point mesh is usually needed. 

\par \textit{Step} 2. With $|\Psi_{i\bf{k}}\rangle$, the projector defined in Eq.~(\ref{Eq20}) can be straightforwardly constructed. 
To this end, the most localized $d$ or $f$ orbital of the correlated atoms in the NAO basis set is used to construct $|W_{I,m}^0\rangle$.

\par \textit{Step} 3. In the DFT+DMFT iteration loop, the frequency-dependent self-energy $\Sigma(i\omega_n)$ is determined by 
the impurity solver at each iteration step. To start with, the initial self-energy is 
set to be equal to the double-counting term, i.e., $\bar{\Sigma}_{ij}(\mathbf{k},i\omega_n)=0$. 
Here the following choice of the double-counting term
\begin{equation}\label{Eq24}
  H_{\mathrm{dc},mm^\prime}^{I, \sigma} = \left[U(n_I-1/2) -1/2 J(n_I-1)\right]\delta_{mm^\prime}
\end{equation}
is used. Here $n_I$ is the total number of correlated electrons associated with the correlated 
atom $I$ and is fixed during the DMFT cycles. In the spirit of reducing the necessity of introducing additional empirical parameters, 
$n_I$ is given by projecting the KS orbitals in the subset $\mathcal{C}$ to the local subspace as
\begin{equation}
 n_I = \sum_{m}\sum_{\mathbf{k}}\frac{1}{N_\mathbf{k}} \sum_{i \in \mathbf{C}} f_{i\mathbf{k}} \tilde{P}_{ii}^{I}\left(\mathbf{k},m m\right) \, ,
 \label{eq:nominal_number}
\end{equation}
where $f_{i\mathbf{k}}$ is the occupation number of KS orbital $\Psi_{i\mathbf{k}}$.
This double-counting scheme is similar to the so-called fixed double-counting~\cite{Pourovskii2007,Haule2014} scheme, which is considered
to be able to improve the stability of DFT+DMFT self-consistency loop~\cite{Haule2014} by fixing the value of $n_I$. 
The difference is that  the nominal number of strongly correlated electrons is specified manually in the fixed double-counting scheme,
whereas in our case this number is determined using Eq.~(\ref{eq:nominal_number}).

\par \textit{Step} 4. Using the projector constructed in \textit{Step} 2, we embed (upfold) the self-energy back to the selected KS space 
subset $\mathcal{C}$, which is expressed as
\begin{equation}\label{Eq23}
  \begin{aligned}
  \bar{\Sigma}_{ij}(\mathbf{k},i\omega_n) =& \sum_{I}\sum_{m m^\prime} \tilde{P}_{ij}^{I}\left(\mathbf{k},m m^\prime\right) \\
    & \left(\Sigma_{mm^\prime}^{I}(i\omega_n)-H_{\mathrm{dc},mm^\prime}^{I}\right) .
  \end{aligned} 
\end{equation}

\par \textit{Step} 5. During the DFT+DMFT self-consistency iteration, the electronic chemical potential needs to be adjusted according to the newly obtained self-energy 
at each iteration, to keep the number of electrons hosted by KS bands in $\mathcal{C}$
\begin{equation}
    N_{\mathcal{C}}^{\mathrm{KS}} = \sum_{\mathbf{k}, i \in \mathcal{C}} \frac{1}{N_{\mathbf{k}}} f_{i\mathbf{k}}
\end{equation}
conserved. Within the DMFT cycle, this condition means that
\begin{equation}\label{Eq25}
  \begin{aligned}
    N_{\mathcal{C}}^{\mathrm{KS}} =& \frac{1}{\beta} \sum_{\omega_n} \sum_{\mathbf{k}, i \in \mathcal{C}} \frac{1}{N_{\mathbf{k}}} \\
    & \left[ \frac{1}{i\omega_n + \mu - \epsilon_{\mathbf{k}} - \bar{\Sigma}(\mathbf{k},i\omega_n) } \right]_{ii} \, ,
  \end{aligned}
\end{equation}
where $\beta$ is again the inverse temperature $1/{k_B T}$. The summation of imaginary frequency $\omega_n$ should run 
from negative infinity to positive infinity. However, in realistic calculations, to save computational cost,
the explicit summation over Matsubara frequency points is only carried within a frequency window $[-\omega_N, \omega_N]$, where
the contribution from the frequency points outside the window is treated approximately. This is enabled by 
making use of the asymptotic behavior of the self-energy, i.e., $\lim_{n \to \infty} \bar{\Sigma}_{ii}(\mathbf{k}, i\omega_n) = \bar{\Sigma}_{ii}(\mathbf{k}, \infty)$, 
where $\bar{\Sigma}_{ii}(\mathbf{k}, \infty)$ is a real value. Then the Eq.~(\ref{Eq25}) is approximated by 
\begin{equation}\label{Eq26}
  \begin{aligned}
    N_{\mathcal{C}}^{\mathrm{KS}} =& \frac{1}{\beta} \sum_{\mathbf{k}, i \in \mathcal{C}} \frac{1}{N_{\mathbf{k}}} \left\{ \sum_{ \omega_n = -\omega_N }^{\omega_N }  \right.  \\
    & \left( \left[ \frac{1}{i\omega_n + \mu - \epsilon_{\mathbf{k}} - \bar{\Sigma} (\mathbf{k}, i\omega_n) } \right]_{ii}  \right. \\
    & \left. -  \left[ \frac{1}{i\omega_n + \mu - \epsilon_{k} - \bar{\Sigma} (\mathbf{k}, \infty) } \right]_{ii} \right)  \\
    & \left. + \frac{1}{1+e^{\beta(\epsilon_{i\mathbf{k}} + \bar{\Sigma}_{ii}(\mathbf{k}, \infty) -\mu)}} \right\} .
  \end{aligned}
\end{equation}
When the chosen cutoff Matsubara frequency $\omega_N$ is high enough so that the asymptotic behavior of the self-energy
is correct, this approximation is of good accuracy.

\par \textit{Step} 6. As all information is secured, the local interacting Green function is constructed from Eq.~(\ref{Eq17}), where, of course, the renormalized projector Eq.~(\ref{Eq20}) is used. 
Under the DFT+DMFT self-consistency condition, Eq.~(\ref{Eq1}) and (\ref{Eq8}), the matrices of the impurity level and 
hybridization function are determined by
\begin{equation}\label{Eq27}
  {\cal E}_{I, m m^{\prime}}^{imp} = -H_{dc,mm^\prime}^{I} + \sum_{\mathbf{k}, i \in \mathcal{C}} \frac{1}{N_{\mathbf{k}}} \tilde{P}_{ij}^{I}\left(\mathbf{k},m m^\prime\right) \varepsilon_{i \mathbf{k}} 
\end{equation}
and 
\begin{equation}\label{Eq28}
  \begin{aligned}
  \Delta_{m m\prime}^{I} \left( i\omega_n \right) = &(i\omega_n + \mu)\delta_{m m^\prime} - {\cal E}^{imp}_{I,m m^\prime} \\ 
  & -\Sigma_{m m^\prime}^{I}\left(i\omega_n\right) - [G^{loc}]^{-1}_{I, m m^\prime} .
  \end{aligned} 
\end{equation}
For some CT-HYB impurity solver, the imaginary time hybridization function is needed. 

\par \textit{Step} 7. Solve the impurity problem with the determined $\mu$, ${\cal E}_{I}^{imp}$, $\Delta^{I} \left( i\omega_n \right)$ and 
the given on-site Coulomb interaction through the impurity solver to obtain the new self-energy and impurity Green function. In this paper,
we use the Kanamori \cite{Kanamori1963,Anisimov2014} form Coulomb interaction in which only the density-density term is included, 
so there are no sign problem in our CTQMC calculations.

\par \textit{Step} 8. Check whether the self-consistency is reached. If the self-consistency is reached, stop the DFT+DMFT calcution,
else return to \textit{Step} 4.

\section{Results and discussion}\label{Sec3}
\subsection{\textit{d}-electron systems}
\par We first benchmark our DFT+DMFT implementation on three prototypical strongly correlated \textit{d}-electron systems -- SrVO$_3$, NiO, and MnO. 
For the DFT part, we carried out GGA calculations using two NAO-based code packages -- FHI-aims \cite{Blum2009} and ABACUS \cite{Li2016}. In FHI-aims, 
the default \textit{tight tier 1} basis set is used for V, Ni, Mn, O, and Sr atoms, and the corresponding cutoff radii of the basis functions
are \SI{6.0}{\angstrom}, \SI{6.0}{\angstrom}, \SI{6.0}{\angstrom}, \SI{6.0}{\angstrom}, and \SI{8.0}{\angstrom}, respectively. 
In ABACUS we use the SG15 optimized norm-conserving Vanderbilt (ONCV) multi-projector pseudo-potentials \cite{Hamann2013, Schlipf2015, Scherpelz2016} 
and the corresponding optimized double-$\zeta$ plus polarization (DZP) atomic basis sets, which comprise $4s2p2d1f$ basis functions with a cutoff radius of \SI{9.0}{Bohr} 
for transition metal atoms and $2s2p1d$ basis functions with  cutoff radii of \SI{7.0}{Bohr} for O atoms and \SI{10.0}{Bohr} for Sr atoms. 
In all DFT calculations, the Perdew-Burke-Ernzerhof (PBE) GGA exchange-correlation functional was used \cite{Perdew1996-prl}. As for the DMFT part, 
we employed three different CT-HYB impurity solvers to solve the single-site impurity problem. These are PACS \footnote{The segment implementation~\cite{Werner2006} 
of the ct-qmc impurity solver is a part of the PACS@sdf package (Package for Analyzing Correlated Systems with Spatial and Dynamical fluctuations).  
PACS@sdf aims at providing an integrated framework for the study of strongly correlated models and 
materials beyond the local approximation of the DMFT~\cite{Georges1996}. 
It takes DMFT as the zero-order approximation and systematically provides non-local corrections to it~\cite{Li2015}.} developed 
by one of the present authors Gang Li, iQIST developed by Huang \textit{et al.} \cite{Huang2015,Huang2017,iQIST}, and the one used in eDMFT \cite{Haule2010} developed by Haule at Rutgers University \cite{Haule2007,CTHYB-Rutgers}, which is referred to as ``Rutgers" below in this paper.

\subsubsection{SrVO$_3$}
\par SrVO$_3$ has a simple cubic perovskite structure without magnetism.   
DFT fails to reproduce the upper and lower Hubbard bands observed in experiments, and this
calls for advanced theoretical and computational techniques~\cite{Kotliar2006}. In the past, this material has been extensively studied both theoretically and experimentally \cite{Pavarini2004, Nekrasov2006, Liebsch2003, Sekiyama2004, Nekrasov2005, Yoshida2005, Eguchi2006, Maiti2001}, 
which provides abundant reference results. Therefore, SrVO$_3$ is an ideal test example for DFT+DMFT implementations \cite{Anisimov2005, Lechermann2006, Amadon2008-dmft, Anisimov2014, Sim2019}.
The V$^{4+}$ cation with only one 3\textit{d} electron is located in the center of the octahedron formed by its six surrouding
axial O$^{2-}$ ligand anions. In the presence of an octahedral crystal field, the five degenerate \textit{d} orbitals 
split into two subsets: Three-fold degenerate t$_{2g}$ orbitals (i.e., $d_{xy}$, $d_{yz}$ and $d_{xz}$), and two-fold degenerate e$_g$ orbitals 
(i.e., $d_{z^2}$ and $d_{x^2-y^2}$). The only one 3\textit{d} electron of V$^{4+}$ occupies the lower-energy t$_{2g}$ orbitals 
with the higher-energy e$_g$ orbitals being empty. As a common practice, we only consider three degenerate t$_{2g}$ orbitals in our DFT+DMFT calculation.

\par In DFT calculations, we use 11$\times$11$\times$11 $\bfk$-point mesh generated by the Monkhorst-Pack method \cite{Monkhorst1976}. 
The Hubbard \textit{U} and Hund \textit{J} parameters are set to be \SI{4.0}{\electronvolt} and \SI{0.65}{\electronvolt}, respectively, 
following the choice in the literature \cite{Lechermann2006, Amadon2008-dmft, Sim2019}. 
The DMFT calculation is carried out at a temperature of \SI{300}{\kelvin}. 
DFT calculations give a group of bands around the Fermi level with substantial 3\textit{d} characterizes, 
which are well separated from other bands \cite{Lechermann2006, Amadon2008-dmft}. We enclose the six (twelve if spin degree of freedom is taken into account) 
bands crossing the Fermi level in the subset of KS bands $\mathcal{C}$. 

\begin{figure*}[htbp]
  \centering
  \includegraphics[width=0.8\linewidth]{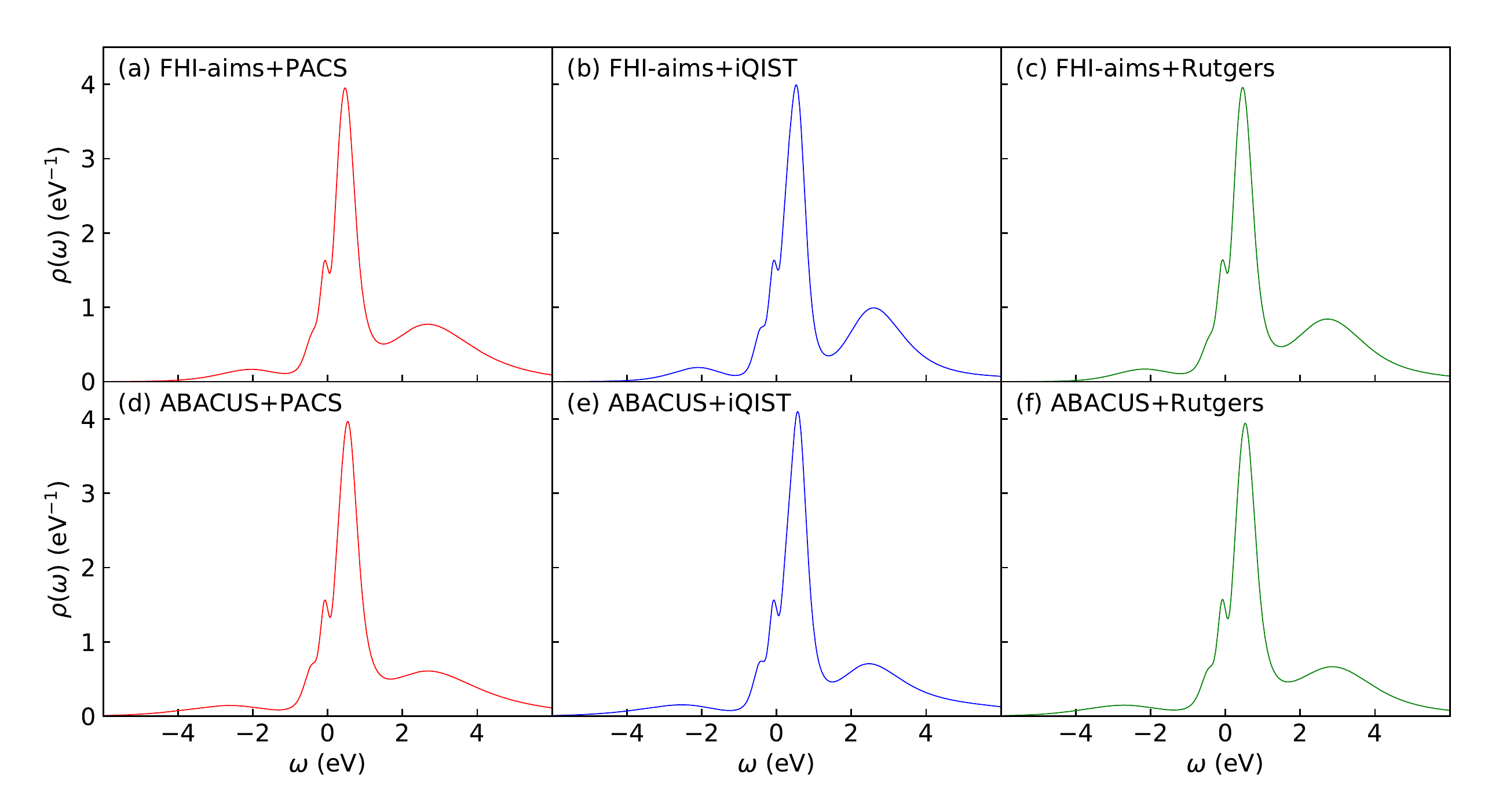}
  \caption{The single-particle energy spectral function of SrVO$_3$ 3\textit{d} electrons given from direct analytical continuation of 
  the impurity Green function.
  The results presented in the six panels are obtained by six computational schemes combining two DFT codes -- FHI-aims (upper panels) and ABACUS (lower panels) 
  and three different impurity solvers -- PACS (left panels), iQIST (middle panels), and Rutgers (right panels).}
  \label{SrVO3}
\end{figure*}

\par One of the powerful strengths of Green-function based DFT+DMFT approach is that it can deliver physically meaningful single particle excitation
spectral functions, as given by the the imaginary part of Green function.  Such spectral functions can be directly measured by photoemission 
and/or inverse photoemission experiments. 
In Fig.~\ref{SrVO3}, the spectral functions of 3\textit{d} electrons of SrVO$_3$ calculated by combining 
two NAO-based DFT packages -- FHI-aims and ABACUS, and three impurity solvers -- PACS, iQIST and Rutgers are presented. 
Despite the considerable differences underlying the implementations of the DFT codes, as well as the DMFT impurity solvers,
the calculated spectral functions are remarkably similar. 
Most importantly, they all successfully reproduce the typical three-peak structure: A quasi-particle 
band located at the Fermi level with the lower and upper Hubbard bands on the two sides, 
arising from the strong electron correlations which traditional LDA and GGA fail to capture.
The main features of the spectral functions given by our DFT+DMFT calculations are in good agreement with 
previous theoretical results except for some small details
such as the exact peak positions and intensities~\cite{Sekiyama2004, Pavarini2004, Nekrasov2005, Anisimov2005, Nekrasov2006, Lechermann2006, Amadon2008-dmft, Anisimov2014, Sim2019}. Our results also correctly describe the strong particle-hole asymmetry of the lower and upper Hubbard band as 
the intensity of the occupied lower Hubbard band is much smaller than the empty upper Hubbard band. This is
the consequence that the V$^{4+}$, with only one 3\textit{d} electron, is far away from half-filling.

\par Similar to previous DFT+DMFT studies \cite{Pavarini2004, Nekrasov2005, Anisimov2005, Nekrasov2006, Lechermann2006, Amadon2008-dmft, Anisimov2014, Sim2019}, 
our spectral function results are in excellent agreement with the experimental
photoemission spectrum \cite{Sekiyama2004}. The first main feature lies in the reproduction of the lower Hubbard
band at around \SI{-2.0}{\electronvolt}, which is the manifestation of the strong correlation among 3\textit{d}
electrons. The second feature, in quantitative agreement with experiment, is that the lower Hubbard band vanishes at $\sim \SI{-1.0}{\electronvolt}$, 
and then the quasiparticle band starts to rise sharply.

\begin{figure*}[htbp]
  \centering
  \includegraphics[width=0.8\linewidth]{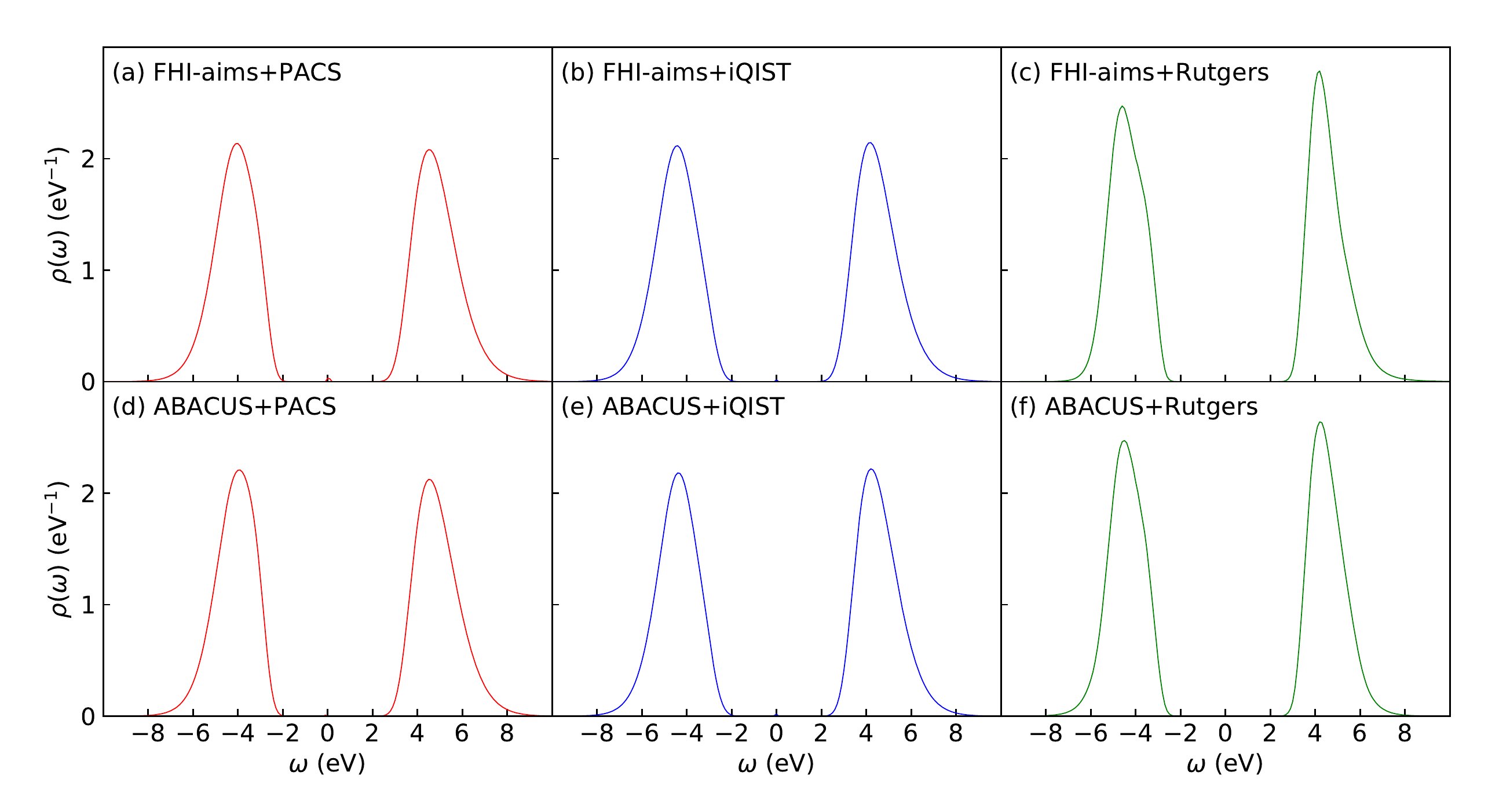}
  \caption{The spectral function of MnO 3\textit{d} electrons obtained from direct analytical continuation of 
  the impurity Green function. 
  Similar to Fig.~\ref{SrVO3}, results presented in the six panels are obtained by 
  two different DFT codes combined with three different impurity solvers. Each panel is labelled by the
  specific combination.} 
  \label{MnO-Aw}
\end{figure*}

\subsubsection{MnO and NiO}
\par Next we test our DFT+DMFT scheme on the late transition metal oxides MnO and NiO. 
Both MnO and NiO crystallize in face-centered cubic (FCC) NaCl-type structures where 3\textit{d} orbitals are split into 
threefold degenerate $t_{2g}$ and twofold degenerate $e_g$ orbitals as in the SrVO$_3$ case. In the MnO system, 
we adopt the parameters of \textit{U}$=$\SI{5.0}{\electronvolt}, \textit{J}$=$\SI{1.0}{\electronvolt}, and 
the temperature $T=$\SI{300}{\kelvin}. For MnO, the KS bands within the energy window of [\SI{-3.0}{\electronvolt}, \SI{3.0}{\electronvolt}] are included in the subset $\mathcal{C}$.
For NiO, another set of parameters with the temperature of \SI{1160}{\kelvin}, \textit{U} of \SI{8.0}{\electronvolt} and 
\textit{J} of \SI{1.0}{\electronvolt} \cite{Anisimov1991, Ren2006, Kunes2007, Sim2019} is used.
The energy window enclosing the subset of KS bands $\mathcal{C}$ is chosen to be [\SI{-2.0}{\electronvolt}, \SI{1.0}{\electronvolt}].
\begin{figure*}[htbp]
  \centering
  \includegraphics[width=0.8\linewidth]{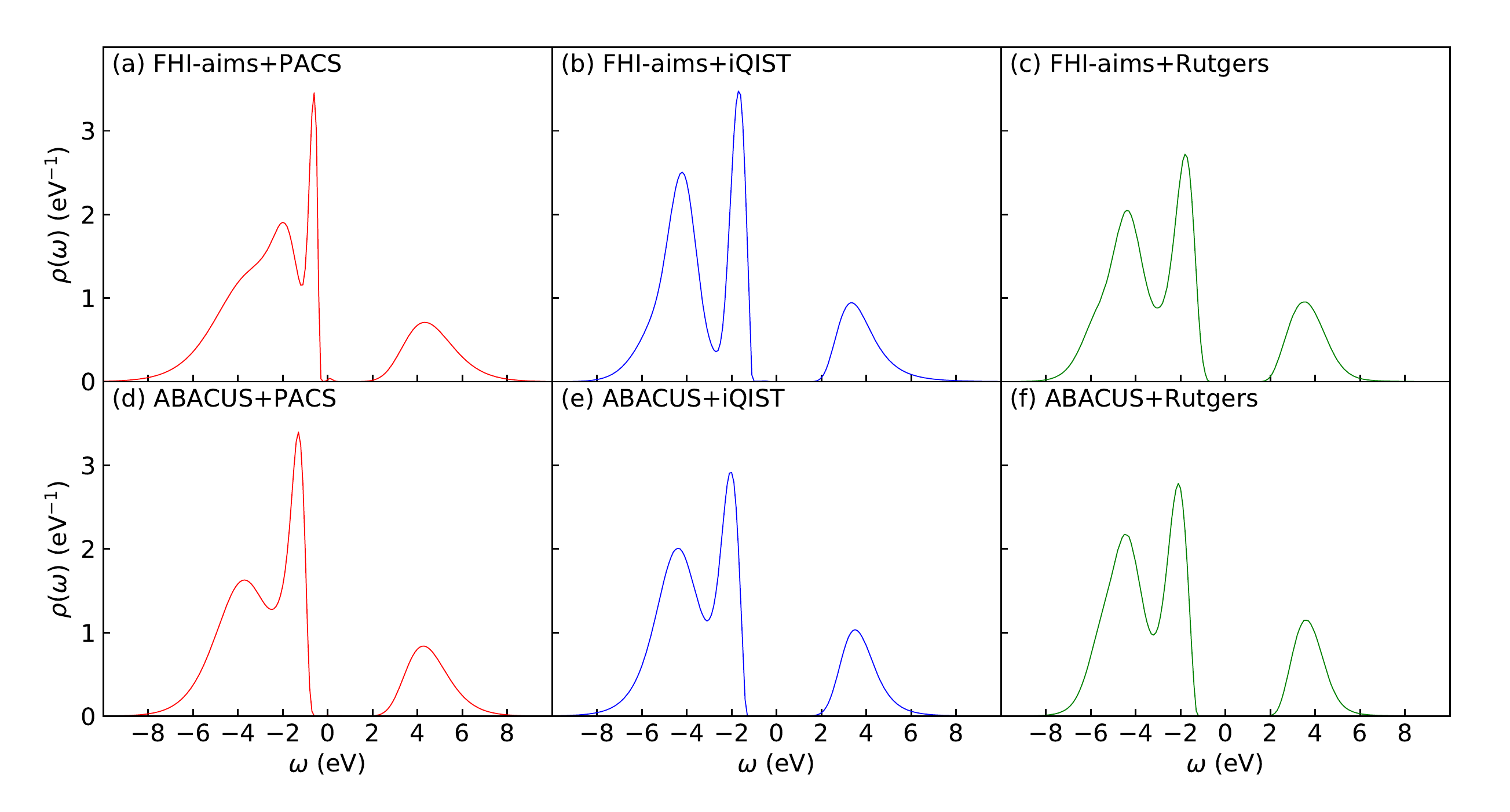}
  \caption{The single-particle spectral function of NiO 3\textit{d} electron calculated from direct analytical continuation of the impurity Green function. 
  The meaning of the six panels is same as Fig.~\ref{SrVO3} and \ref{MnO-Aw}.}
  \label{NiO-Aw}
\end{figure*}

\begin{figure*}[htbp]
  \centering

  \begin{minipage}[htb]{0.4\linewidth}
    \centering
    \includegraphics[width=\linewidth]{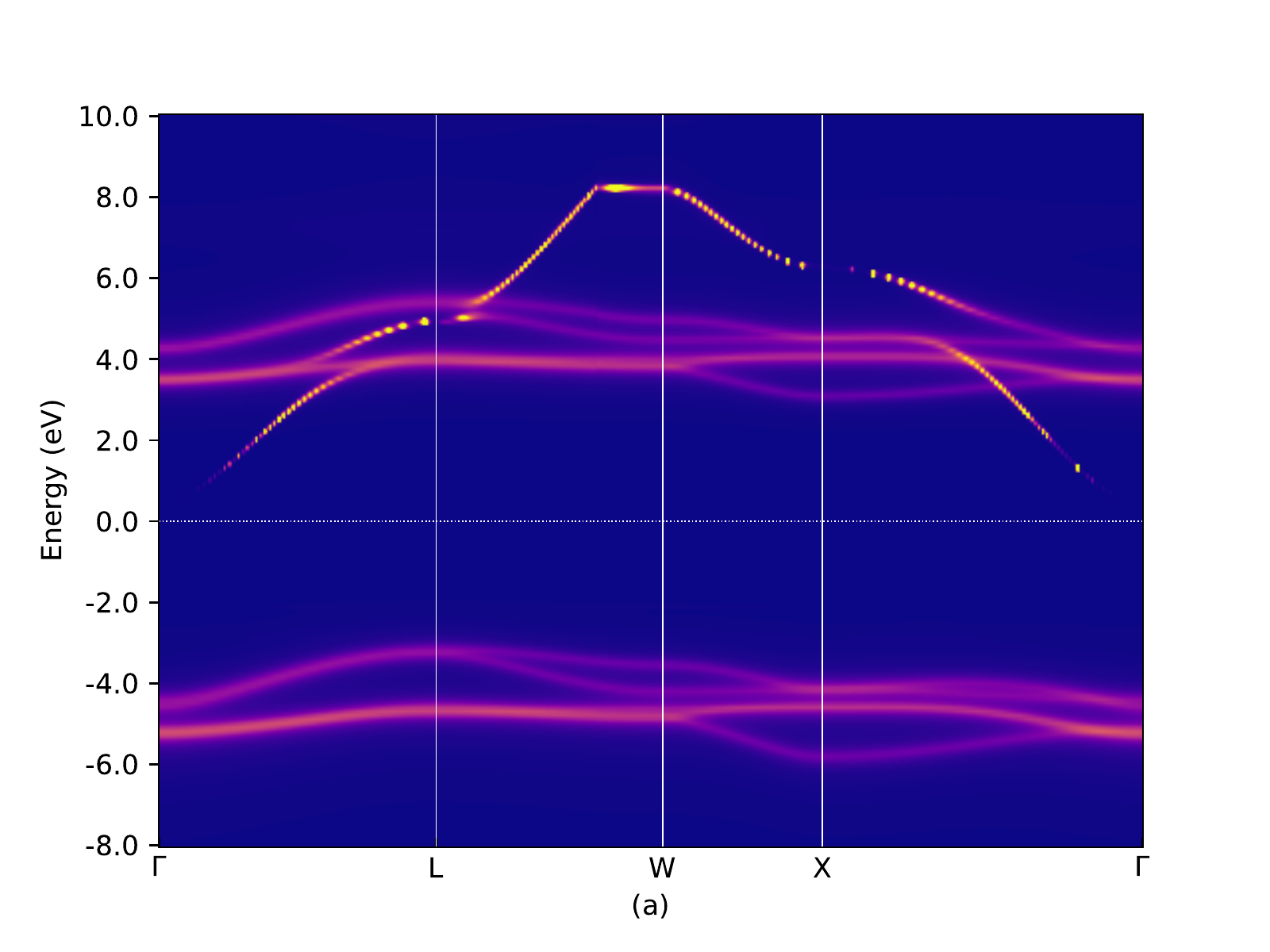}
  \end{minipage}
  \begin{minipage}[htb]{0.4\linewidth}
    \centering
    \includegraphics[width=\linewidth]{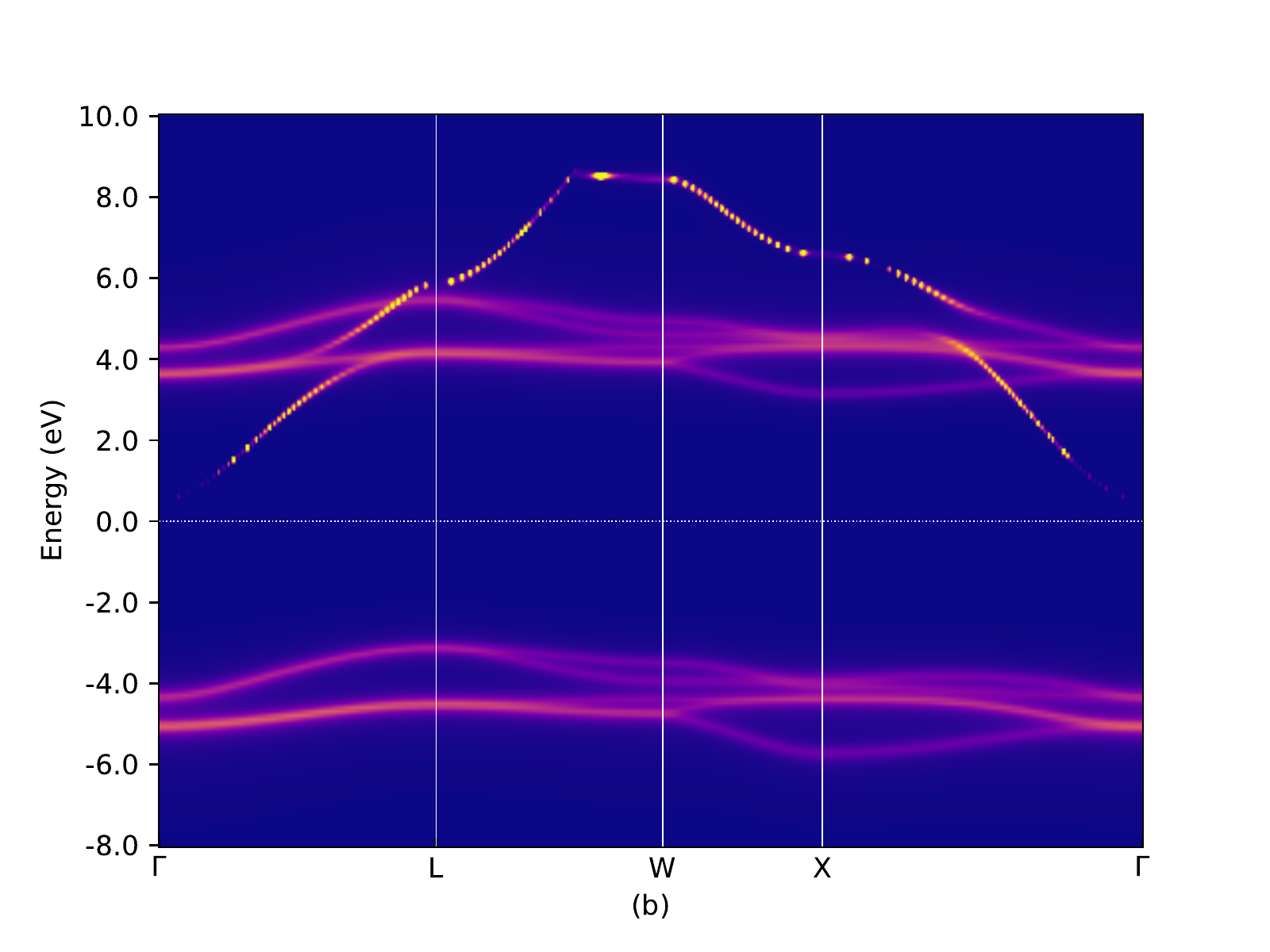}
  \end{minipage}

  \caption{The $\bfk$-resolved spectral function $A(\mathbf{k},\omega)$ of MnO, obtained by (a) FHI-aims+Rutgers and (b) ABACUS+Rutgers,
  respectively.}
  \label{MnO-Awk}
\end{figure*}

\begin{figure*}[htbp]
  \centering
  \begin{minipage}[htb]{0.4\linewidth}
    \centering
    \includegraphics[width=\linewidth]{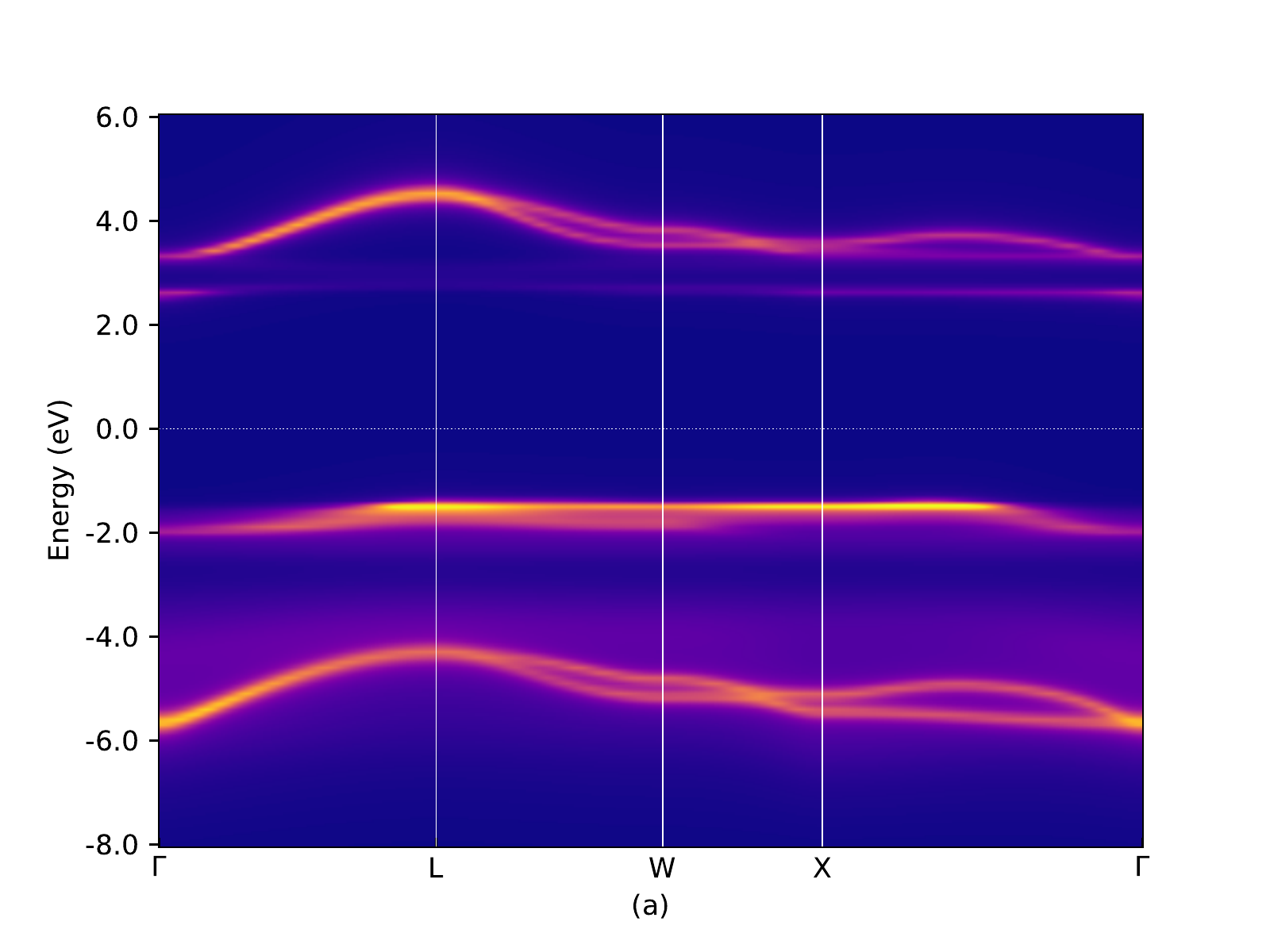}
  \end{minipage}
  \begin{minipage}[htb]{0.4\linewidth}
    \centering
    \includegraphics[width=\linewidth]{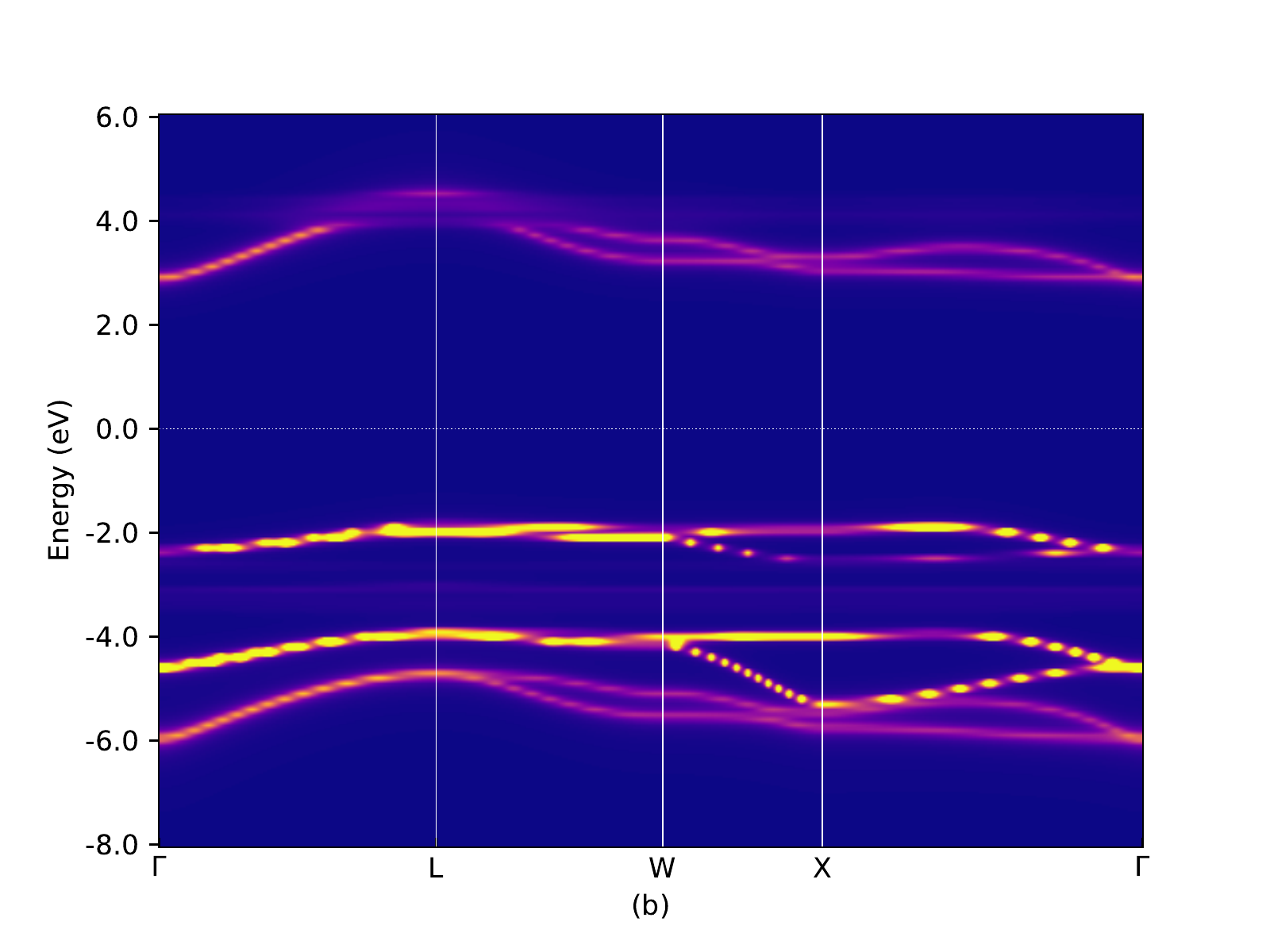}
  \end{minipage}
  \caption{The k-resolved spectral function $A(\mathbf{k},\omega)$ of NiO, obtained by (a) FHI-aims+Rutgers and (b) ABACUS+Rutgers, respectively.}
  \label{NiO-Awk}
\end{figure*}

\par MnO and NiO are classical examples that clearly demonstrate the failure of traditional band theory, 
e.g., LDA and GGA. LDA and GGA predict MnO and NiO to be metallic while they are wide-gap insulators experimentally. 
Even though a gap will appear if the spin-symmetry-broken antiferromagnetic state is considered,
the calculated gap is however still one order of magnitude smaller than the experimental values. 
By merging the Hubbard model and 
\textit{ab}-\textit{initio} DFT, static DFT+\textit{U} \cite{Anisimov1991, Anisimov1993, Anisimov1997_LDAU} and the 
more advanced dynamical DFT+DMFT \cite{Ren2006,Kunes2007} successfully open the gaps in those prototypical
transitional metal oxides. 
Then obtained band gaps are in quantitative agreement with experiments, if proper parameters are used.

\par We then apply our DFT+DMFT implementation to both MnO and NiO, and the obtained spectral functions are presented
in Fig.~\ref{MnO-Aw} and Fig.~\ref{NiO-Aw}, respectively. 
Again in each figure, we present six panels containing results obtained by combining two NAO-based DFT codes and three DMFT impurity solvers.
The most prominent feature in the DFT+DMFT spectra of both MnO and NiO is that a sizeable gap is opened up, arising apparently from
the strong Coulomb interactions whose effect is properly captured by DMFT.
One can further see from Fig.~\ref{MnO-Aw}  that the intensity of the lower and upper Hubbard band of Mn 3\textit{d} electrons
are nearly the same. This result is consistent with the nominally half-filling 3\textit{d} orbitals of Mn given by chemical analysis, 
which is confirmed by the fact that the number of 3\textit{d} electrons of Mn 
given by $n=-\sum_{m} G^{imp}_{mm}(\beta)$ is 5.
Similar to the SrVO$_3$ case, the results presented in all six panels are very close, except that the peaks given by the Rutgers impurity
solver is slightly more pronounced. 

\par For NiO, the number of nominal $3d$ electrons is approximately 8, and DFT+DMFT calculations lead to a splitting of the $3d$ bands, 
with fully occupied $t_{2g}$ orbitals
and half-filled $e_g$ orbitals. Our results in general agree well with previous DFT+DMFT results reported in the literature \cite{Ren2006, Kunes2007}. 
However, Fig.~\ref{NiO-Aw} also reveals that both the DFT codes and the impurity solvers have certain influence on the obtained DFT+DMFT
energy spectra. Although there are no qualitative differences, the shape and width of the left peak, and the depth of the dip between the first and middle peaks show noticeable quantitative
differences. It is not entirely clear to us yet what factors caused such differences. We note that, however, NiO is a prototypical charge transfer
insulator \cite{Zaanen1985}, and the competition between the strong Coulomb interactions among the Ni $3d$ electrons and the hybridization 
between the Ni $3d$ and O $2p$ electrons governs its underlying physics. Thus a complete DFT+DMFT treatment of NiO should also include O $2p$ electrons
in the game. With the O $2p$ excluded, the DFT+DMFT calculations are probably more sensitive to the details in the band structures and numerical techniques
behind the impurity solvers.

\par Figure~\ref{MnO-Aw} and \ref{NiO-Aw} only contain the spectral information of 3\textit{d} electrons. 
Our DFT+DMFT implementation also allows for calculating the $\bfk$-resolved spectral function
in the KS orbital space, as determined by 
\begin{equation}\label{Eq-Awk}
    A(\mathbf{k},\omega) = -\frac{1}{\pi}\sum_{i \in \mathcal{C}} \operatorname{Im}\left[(\omega + \mu - 
    \epsilon({\mathbf{k}}))I  - \bar{\Sigma}( \mathbf{k}, \omega) \right]^{-1}_{ii},
\end{equation}
where the real-frequency self-energy is evaluated by analytical continuation of the imaginary frequency-self-energy 
through the maximum entropy formalism in Ref.~\cite{Kraberger2017}. The corresponding results for MnO and NiO are presented in 
Figs.~\ref{MnO-Awk} and \ref{NiO-Awk}, respectively. Now for simplicity only results obtained using the two DFT codes combined with 
the Rutgers impurity solver are presented. The agreement between our theoretical gaps and experimental gaps \cite{Messick1972, Sawatzky1984, Hufner1984} 
is rather  satisfactory. The two $\bfk$-resolved spectral functions of MnO is nearly identical whereas again noticeable 
differences, in particular regarding the spectral weights of certain bands, exist for NiO. 
Nevertheless the main features, e.g., the energy positions and dispersion of the bands, are reasonably similar 
in Fig.~\ref{NiO-Awk}(a) and \ref{NiO-Awk}(b).

\par
\begin{figure}[htbp]
  \centering
  \includegraphics[width=\linewidth]{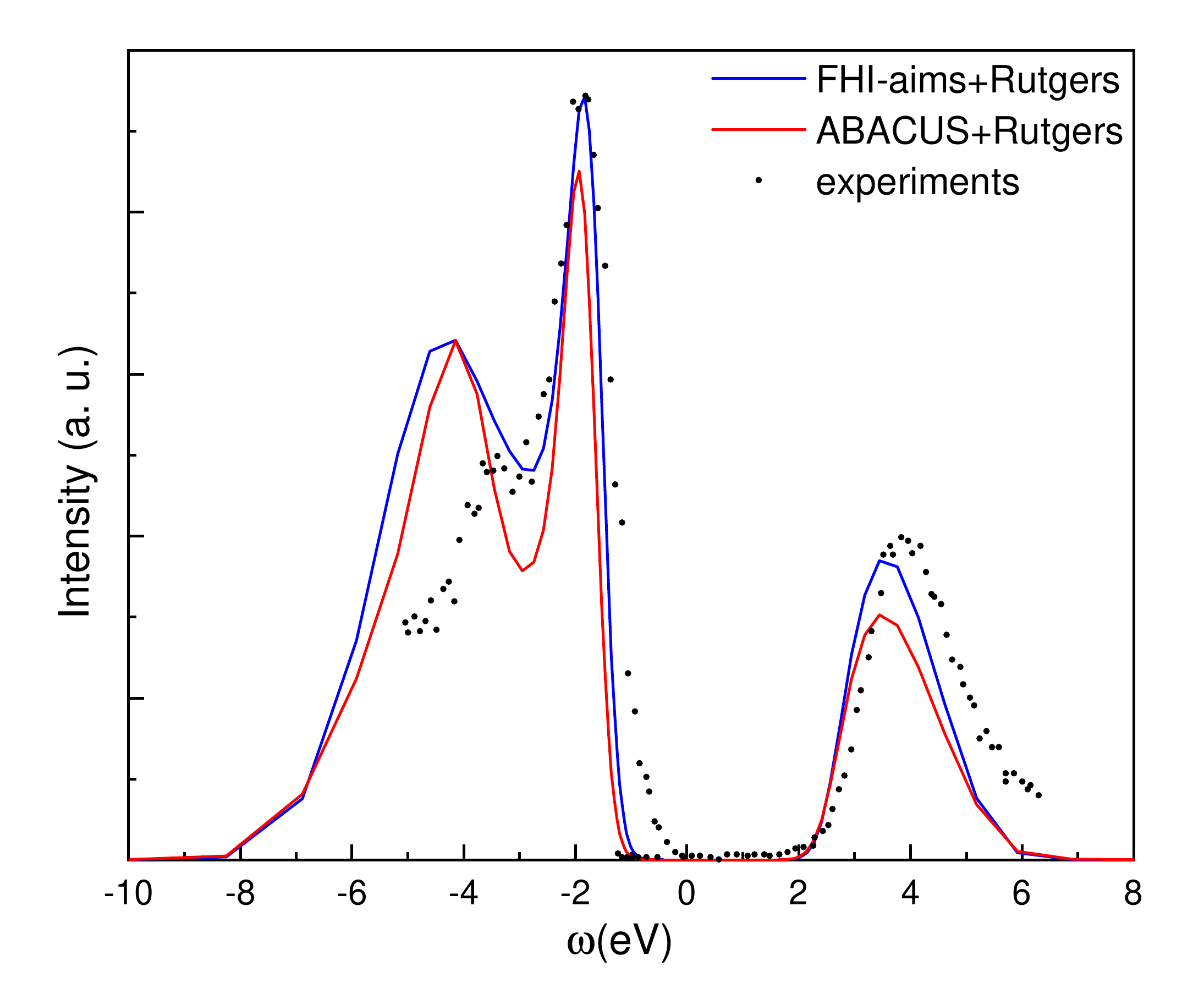}
  \caption{Comparison of theoretical and experimental spectral functions of NiO. The theoretical results are given by the imaginary part 
  of the impurity Green function on the real frequency axis, analytically continued from the Matsubara impurity Green function. 
  The experimental data are obtained from XPS+BIS measurements, taken from Ref.~\cite{Sawatzky1984}.}
  \label{NiO-Aw-exp}
\end{figure}

 NiO has been extensively studied experimentally, and reliable experimental photoemission spectra are available. 
In Fig.~\ref{NiO-Aw-exp}, we directly compare our DFT+DMFT spectrum of NiO  with the experimental spectrum 
of Sawatzky and Allen \cite{Sawatzky1984} measured by x-ray-photoemission (XPS) and 
bremsstrahlung-isochromat-spectroscopy (BIS) techniques. Again, the two sets of theoretical spectra 
are obtained using FHI-aims and ABACUS codes interfaced with the Rutgers impurity solver, respectively. 
For the spectrum below the Fermi level, corresponding to the (negative of) energy cost for 
removing a particle from the occupied levels, 
we reproduce the sharp peak around \SI{-2.0}{\electronvolt} and a local minimum at \SI{-3.0}{\electronvolt} given by XPS. 
For the spectrum above the Fermi level, corresponding to the (negative of) energy cost for adding a particle 
to the system, our curves match perfectly with the BIS result. Although small deviations of the shoulder peak at 
approximately $\SI{-3.5}{\electronvolt}$ are visible, our DFT+DMFT spectrum, in general, agrees well 
with previous theoretical work \cite{Ren2006} and the experiment data\cite{Sawatzky1984}. 

\par In ABACUS, the pseudopotential method is used and only the valence electrons are explicitly included.
Furthermore, as given in the beginning of this section, the spatial cutoffs of NAOs used in ABACUS are much smaller than those 
used in FHI-aims. By contrast, in FHI-aims full potential is used with all electrons being included 
in DFT calculations. Although these two NAO schemes are rather different, we get nearly identical DFT+DMFT results of SrVO$_3$ and MnO and 
reasonably similar results for NiO. It proves the efficacy and robustness of our DFT+DMFT formalism and implementation. We expect 
that the DFT+DMFT formalism presented in this paper should work for other NAO-based DFT codes as well. 

\subsection{\textit{f}-electron systems}
\par The \textit{f}-electron systems, including 4\textit{f} lanthanides and 5\textit{f} actinides, are an important class of 
strongly correlated materials characterized by partially filled \textit{f}-type orbitals. 
These systems exhibit a variety of exotic phenomena, such as the heavy-fermion behavior, metal-insulator transition, 
Kondo physics, volume collapses accompanying phase transitions, etc.  
It has been generally accepted that traditional DFT calculations based on static mean-field-type approximations such as LDA and GGAs
do not provide adequate accuracy for describing these physical scenarios. DFT+DMFT 
has proved to be a powerful approach to tackle these systems and achieved remarkable successes in the last two decades. 
In this section we test our DFT+DMFT implementation on several typical \textit{f}-electron systems, 
including 4\textit{f} systems like $\alpha$ and $\gamma$ cerium (Ce) metal and Ce$_2$O$_3$, and 5\textit{f} systems like PuO$_2$ and Pu$_2$O$_3$. 

\par DFT calculations are carried out using FHI-aims with the default \textit{tight tier1} basis sets. 
For these basis sets, the cutoff radii of Ce, Pu, and O elements are \SI{7}{\angstrom}, \SI{6.0}{\angstrom}, 
and \SI{6.0}{\angstrom}, respectively. Analogous to the case of $3d$-electron systems, the GGA-PBE functional \cite{Perdew1996-prl}  is
used in the DFT calculations and  11$\times$11$\times$11 Monkhorst-Pack 
$\bfk$-point mesh \cite{Monkhorst1976} is used for Brillouin-zone integration, which is deemed dense enough to obtain accurate DFT band structures. 
Unlike the practice in previous sections where results obtained using three DMFT impurity solvers are shown for comparison, here
for simplicity only the results obtained using the Rutgers solver will be presented.

\subsubsection{Ce metal}
\par Despite its simple FCC structure and the fact that only one 4\textit{f} valence electron is present per Ce atom, Ce exhibits 
spectacular physical properties, which has attracted considerable research interest over the last
forty decades~\cite{Grioni1997,Weschke1991,McMahan1998,Held2001,McMahan2003,Amadon2006,Lipp2012,Anisimov2014,Bieder2014,Casadei2016}. 
At low temperature and ambient pressure, Ce crystalizes in the $\alpha$ phase (smaller volume), where the system is paramagnetic 
and shows Pauli-like magnetic susceptibility without forming local magnetic moments. At high temperature, 
Ce transforms into the $\gamma$ phase (larger volume), which instead carries local magnetic moments and exhibits Curie-Weiss 
behavior of magnetic susceptibility. When increasing pressure or decreasing temperature, Ce undergoes the famous 
isostructural $\gamma$-$\alpha$ phase transition accompanied by a 14\% volume collapse and a drastic change of magnetic 
properties \cite{McMahan1998,Held2001,McMahan2003,Amadon2006,Lipp2012,Anisimov2014,Bieder2014,Casadei2016}. 
A theoretical understanding of this behavior poses a great challenge to condensed matter physics and has motivated lots of experimental 
and theoretical investigations. From the perspective of first-principles computations, standard DFT calculations are unable to give a
proper description of this phase transition, and the approaches going beyond conventional DFT are required.

\par In this paper, the crystal structures of $\alpha$ and $\gamma$ Ce phases are discerned by different volumes, i.e., \SI{29}{\angstrom}$^3$ 
for $\alpha$ and \SI{34}{\angstrom}$^3$ for $\gamma$ phase, 
respectively, which covers the volume ranges during the $\alpha$-$\gamma$ transition \cite{Held2001}. 
The chosen energy window for KS subset $\mathcal{C}$ is [\SI{-5.0}{\electronvolt}, \SI{5.0}{\electronvolt}].
The Coulomb interaction parameter \textit{U} given by constrained DFT is \SI{6.0}{\electronvolt} \cite{Anisimov2014,Bieder2014}. 
Here we used the value of \textit{U} of \SI{6.0}{\electronvolt} and Hund parameter \textit{J} of \SI{0.5}{\electronvolt} \cite{Amadon2006,Bieder2014}. 
DFT+DMFT calculations are done at the temperature of \SI{800}{\kelvin}. 

\begin{figure}[htbp]
  \centering
  \includegraphics[width=\linewidth]{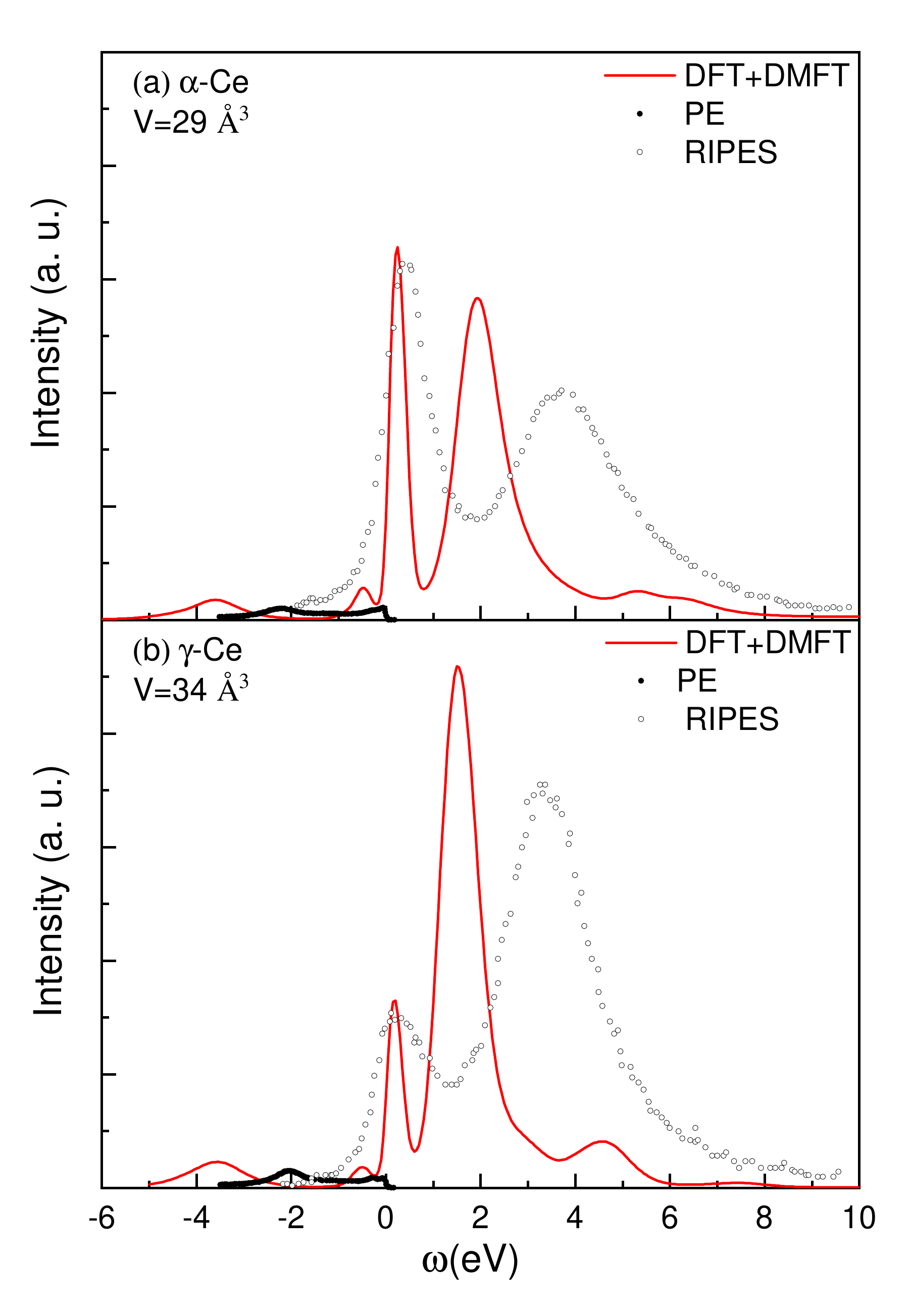}
  \caption{Comparison of theoretical and experimental spectral functions of $\alpha$ and $\gamma$ Ce. 
  The DFT+DMFT calculation is carried out by FHI-aims interfaced with the Rutgers impurity solver and
  the corresponding spectral results are given by direct analytical continuation of the impurity Green function.
  The resonant inverse photoemission spectroscopies (RIPES) are taken from Ref.~\cite{Grioni1997}, and the photoemission (PE) spectra are
  taken from Ref.~\cite{Weschke1991}.}
  \label{Ce}
\end{figure}

\par Our DFT+DMFT spectra for Ce are shown in Fig.~\ref{Ce}. First of all, the energy spectra of both phases show the characteristic 
three-peak structures of strongly correlated metals. One can also see that, going from the $\alpha$ phase to the $\gamma$ phase,  
the central quasiparticle peak gets substantially suppressed,
and its spectral weights transfer to the lower and upper Hubbard bands . 
Such results are in agreement with experimental observations \cite{Grioni1997,Weschke1991} and previous DFT+DMFT calculations for Ce
\cite{Held2001,McMahan2003,Amadon2006,Bieder2014}. The relative strengths of 
the quasi-particle peak and the upper Hubbard peak of both $\alpha$ and $\gamma$ phases are also captured qualitatively.
However, a direct comparison of the calculated spectra with the experimental spectra reveals that the energy separation between
the quasi-particle peak and upper Hubbard band is underestimated in our calculations, while that between
the lower Hubbard band and quasi-particle peak is overestimated. Compared to previous DFT+DMFT calculations, 
the location of the lower Hubbard peak of our results is close to
that reported in the literature \cite{Held2001,Bieder2014} where the maximum of lower Hubbard peaks is 
around -4.0$\sim$\SI{-3.0}{\electronvolt}. 
Concerning the energy positions of the upper Hubbard band, we found that the results reported in the literature also vary quite a bit,
with those reported in Ref.~\cite{Bieder2014} agreeing best with experiment. Our results are close to 
those reported in Ref.~\cite{Amadon2006}, where the upper Hubbard peaks are located at around \SI{2.5}{\electronvolt}.
In general, our DFT+DMFT spectrums are in quantitative agreement with experimental and previous theoretical results.
The discrepancies between our results and experimental and previous theoretical results indicate that
details of the DFT calculations, the definition of the projector, as well as the numerical implementation of the impurity solvers 
will still have appreciable influence on the outcomes of DFT+DMFT calculations. Further investigations along these lines are still needed.

\subsubsection{Ce$_2$O$_3$}
\par Ce$_2$O$_3$ is a Mott insulator with a gap of about \SI{2.4}{\electronvolt} \cite{Prokofiev1996, Pourovskii2007} 
arising from the strong correlation among 4\textit{f} electrons. This system has been calculated using a variety of approaches
such as DFT+\textit{U}, GW \cite{Jiang2009}, and DFT+DMFT \cite{Pourovskii2007, Amadon2012}.
Ce$_2$O$_3$ crystalizes in the hexagonal lattice 
with space group $P\bar{3}m1$. The experimental lattice parameters  $a=b=\SI{3.891}{\angstrom}$, $c=\SI{6.059}{\angstrom}$ \cite{Barnighausen1985}
are used in our calculations. The energy window for KS subset $\mathcal{C}$ is [\SI{-3.0}{\electronvolt}, \SI{2.0}{\electronvolt}].
Constrained DFT calculation predicted the parameter \textit{U} varying from 5.5 to \SI{8.0}{\electronvolt}, and we adopt \textit{U} 
as \SI{6.5}{\electronvolt} and \textit{J} as \SI{0.5}{\electronvolt}. The temperature is set to be \SI{300}{\kelvin}.

\begin{figure}[htbp]
  \centering
  \includegraphics[width=\linewidth]{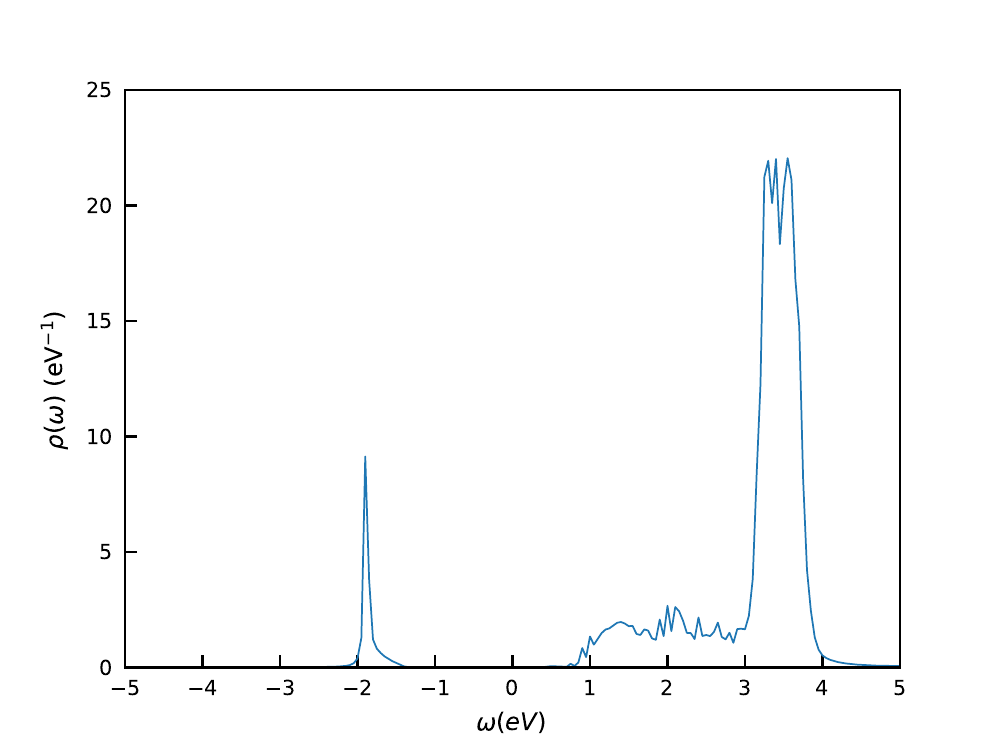}
  \caption{DOS of Ce$_2$O$_3$. The DFT+DMFT calculation is carried out by combination FHI-aims and impurity solver Rutgers.}
  \label{Ce2O3-Aw}
\end{figure}

\par Here we calculate the total density of states (TDOS) of Ce$_2$O$_3$ through 
\begin{equation}\label{Eq-TDOS}
  \rho (\omega) = \frac{1}{N_{\mathbf{k}}} \sum_{\mathbf{k}} A(\mathbf{k},\omega) ,
\end{equation} 
where $A(\mathbf{k},\omega)$ is defined in Eq.~(\ref{Eq-Awk}).
The result is shown in Fig.~\ref{Ce2O3-Aw}, our DFT+DMFT calculation gives a band gap of about \SI{2.5}{\electronvolt}, which agrees
well with the experimental and previous theoretical results \cite{Prokofiev1996, Pourovskii2007, Jiang2009, Amadon2012}. 
For example, all predict that the lower Hubbard peaks are sharp and narrow whereas the upper Hubbard peaks are strong and wide. 
The interesting feature about the upper Hubbard part is that it emerges as a low plateau between 1 and 3 eV above the Fermi level, 
continued by a pronounced peak between 3 and 4 eV.
Note that the peaks mainly composed of O \textit{p} bands sitting below the lower Hubbard peaks does not show up here due to
the fact that those O \textit{p} characteristic bands are not included in the subset $\mathcal{C}$ in the DFT+DMFT calculations
whereas they are retained in Refs.~\cite{Pourovskii2007, Jiang2009, Amadon2012}.

\subsubsection{ Pu$_2$O$_3$ and PuO$_2$}
\par Plutonium (Pu)-based oxides such as Pu$_2$O$_3$ and PuO$_2$ are essential for the fuel components in 
current nuclear reactors, as well as transmutation of the minor actinides from spent nuclear fuels \cite{Carbajo2001,Claparede2017}.
A clear understanding of the physio-chemical properties of plutonium-based oxides is of key importance for the safe operation 
and development of nuclear reactor systems and nuclear waste reprocessing \cite{Arima2005,Petit2010}, and the 
correct description of the oxidation and reduction processes \cite{Moore2017,Li2021}. The physical, chemical, and 
thermodynamical properties of Pu-based oxides such as the chemical bonding and electronic structure 
are intimately related to the states of strongly correlated 5\textit{f} electrons. Conversely, the relative tendency of 
delocalization versus localization of strongly correlated 5\textit{f} electrons is extremely sensitive to the physical 
and chemical environment of the Pu atom \cite{Moore2009}. The description of complicated behaviors of 5\textit{f} electrons, 
e.g., whether they are settled in the delocalized or localized states, or in the intermediate regime, is out of 
the reach of standard DFT. In contrast, DFT+DMFT is becoming a promising approach that facilitates an in-depth understanding of
this type of materials.

\par In this paper, we look at two plutonium oxides, i.e., Pu$_2$O$_3$ in its $\beta$ phase \cite{Wulff1988} (simply called Pu$_2$O$_3$ below) 
and PuO$_2$ \cite{Belin2004} with our DFT+DMFT implementation. We use the Hubbard parameter \textit{U} of \SI{4.0}{\electronvolt} and the
Hund parameter \textit{J} of \SI{0.5}{\electronvolt} for Pu$_2$O$_3$. As for PuO$_2$, \textit{U} and \textit{J} are chosen to be \SI{5.0}{\electronvolt}
and \SI{0.6}{\electronvolt}. The electronic temperatures for both systems are set at \SI{300}{\kelvin}. 
The chosen energy windows for the KS subsets $\mathcal{C}$ are [\SI{-3.0}{\electronvolt}, \SI{2.0}{\electronvolt}] for Pu$_2$O$_3$, and 
[\SI{-3.0}{\electronvolt}, \SI{3.0}{\electronvolt}] for PuO$_2$.

\par
\begin{figure}[htbp]
  \centering
  \includegraphics[width=\linewidth]{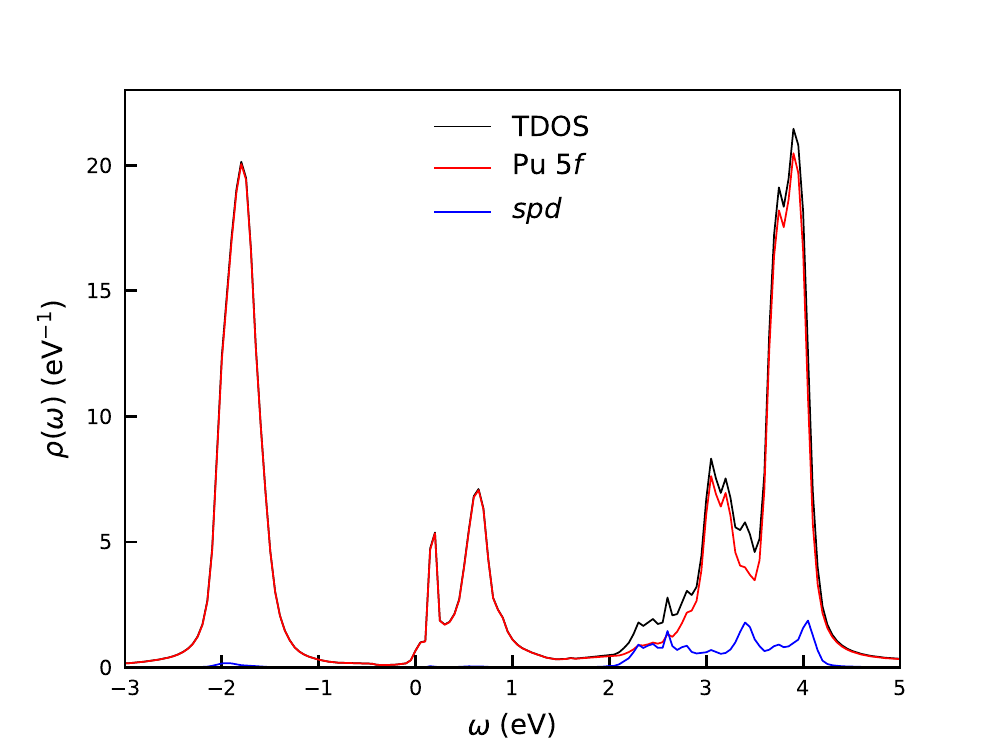}
  \caption{TDOS and PDOS of Pu$_2$O$_3$. The TDOS is given by Eq.~(\ref{Eq-TDOS}). The PDOS of Pu 5\textit{f} electrons is evaluated through Eq.~(\ref{Eq-PDOS}). 
  The PDOS for \textit{spd} electrons is obtained by subtracting the PDOS of Pu 5\textit{f} electrons from the TDOS.
  The DFT+DMFT calculation is carried out by combining FHI-aims and the Rutgers impurity solver.}
  \label{Pu2O3-DOS}
\end{figure}

 We calculated DFT+DMFT TDOS of Pu$_2$O$_3$ through the method introduced in the case of Ce$_2$O$_3$. The TDOS here contains 
contributions from both the correlated orbitals and the rest (here $spd$) orbitals.  The projected density of states (PDOS) 
belonging to correlated ($5f$) orbitals, is evaluated by
\begin{equation}\label{Eq-PDOS}
 \begin{aligned}
   \rho_f (\omega) = & \sum_m \frac{1}{N_{\mathbf{k}}} \sum_{\mathbf{k}} -\frac{1}{\pi}\operatorname{Im} \left\{ \tilde{P}^I_{ij}( \bm{k}, m m) \right.\\
   & \left. \left[(\omega + \mu - \epsilon({\mathbf{k}}))I - \bar{\Sigma}( \mathbf{k}, \omega) \right]^{-1}_{ij} \right\} \, 
 \end{aligned}
\end{equation}
whereas the PDOS for the $spd$ orbitals is obtained by taking the difference between the TDOS and $\rho_f (\omega)$.
The results are depicted in Fig.~\ref{Pu2O3-DOS}. Figure~\ref{Pu2O3-DOS} indicates that the band gap of Pu$_2$O$_3$ as determined by DFT+DMFT is about 
\SI{1.7}{\electronvolt}, which is in good agreement with the experimental bands gap of \SI{1.8}{\electronvolt} \cite{McNeilly1964,Amadon2012}.
Additionally, the occupation analysis yields an average occupation number \textit{n}$_f$=5.0 for Pu 5\textit{f} electrons through $n=-\sum_{m} G^{imp}_{mm}(\beta)$, 
which is consistent with the chemical environment of Pu$^{3+}$ in Pu$_2$O$_3$.

\par
\begin{figure}[htbp]
  \centering
  \includegraphics[width=\linewidth]{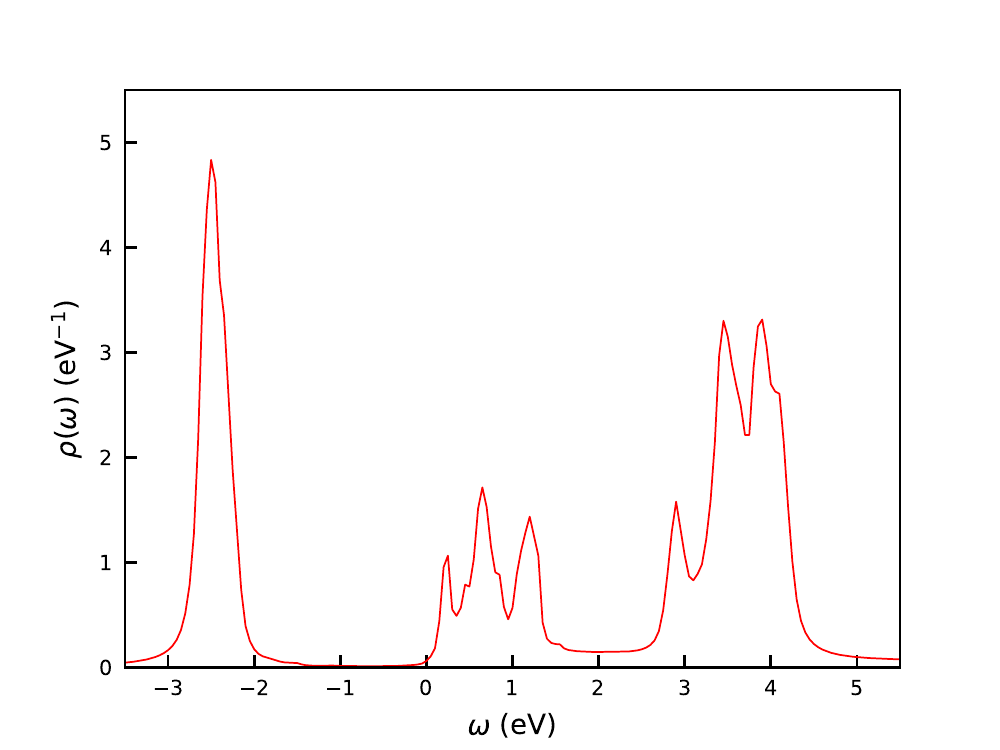}
  \caption{DFT+DMFT TDOS of PuO$_2$. The DFT+DMFT calculation is carried out by a combination FHI-aims and the
  Rutgers impurity solver. }
  \label{PuO2-DOS}
\end{figure}

 In Fig.~\ref{PuO2-DOS}, the DFT+DMFT TDOS of PuO$_2$ is plotted. The bands gap is predicted to be about \SI{2.5}{\electronvolt}, 
which is in good agreement with previous theoretical work \cite{Kolorenc2015} and the experimental band gap of 
\SI{2.8}{\electronvolt} \cite{MarkMcCleskey2013}. In PuO$_2$, the peak of occupied Pu 5\textit{f}
electrons is noticeably sharper than that of Pu$_2$O$_3$, which indicates that the Pu 5\textit{f} electrons in PuO$_2$
are likely more localized than the case of Pu$_2$O$_3$. 
This localization picture is in agreement with the bigger gaps in PuO$_2$. The occupation analysis 
gives an average occupation number of \textit{n}$_f$=4.0 for Pu 5\textit{f} electrons,
which agrees with the  chemical state of Pu$^{4+}$ in PuO$_2$.

\section{Summary}\label{Sec4}
\par In summary, we developed and implemented a formalism that allows us to carry out DFT+DMFT calculations within the 
NAO basis set framework. For transition metal compounds and $f$-electron systems, the most localized $d$ or $f$-type NAO
is used to define the local correlated subspace. Such a choice is physically intuitive and implementationally convenient for NAO basis sets.
Following what is usually done in the literature, only a subset of the KS bands $\mathcal{C}$ out of the full KS space around the Fermi energy,
which hosts the majority of electrons with strong correlations, are enclosed in the definition of the projector. Our projector scheme
is mathematically equivalent to the projective-Wannier-function approach adopted in Ref. \cite{Anisimov2005}. 
We implemented such a DFT+DMFT formalism by interfacing two NAO-based DFT codes, i.e., FHI-aims \cite{Blum2009} and
ABACUS \cite{Li2016} with three DMFT quantum impurity solvers, i.e., PACS, iQIST \cite{Huang2015,Huang2017,iQIST},
and the Rutgers impurity solver \cite{Haule2010,Haule2007,CTHYB-Rutgers}. In particular, interfacing with the all-electron
FHI-aims code allows one to study all types of strongly correlated materials over the periodic table. 

\par Our DFT+DMFT formalism and implementation are testified for three typical series of 
strongly correlated materials, namely the 3\textit{d} transitional metal compounds SrVO$_3$, MnO and NiO, 4\textit{f}
materials including the Ce metal and Ce$_2$O$_3$, as well as the 5\textit{f} actinides Pu$_2$O$_3$ and 
PuO$_2$. For SrVO$_3$ and MnO, our calculated one-electron removal and addition spectra are in good agreement 
with previously reported DFT+DMFT results and experimental data. Furthermore, the calculated results are rather robust against the use of
different DFT codes or different impurity solvers. For NiO, the obtained DFT+DMFT results show noticeable dependence on
the chosen DFT codes and/or impurity solvers,  although the key spectroscopic features are captured by all calculations. For $f$-electron
systems, the characteristic three-peak structures are obtained for the Ce metal, whereas for the correlated insulators, the obtained DFT+DMFT band gaps
are in good agreement with experiments. However, there remain issues calling for further investigations,
like the energy separation of the quasiparticle peak and the upper Hubbard band for the Ce metal. 

\par Admittedly, our scheme is still in its infancy, and at the quantitative level there are still issue to be sorted out. 
Further improvement would be necessary for more reliable descriptions of charge-transfer-type Mott insulators, and the intricate lanthanides and actinides.
However, we consider that our attempt of developing an infrastructure that merge NAO-based DFT codes and DMFT-based techniques is rather
rewarding.  This will not only enable standard DFT+DMFT calculations within the NAO basis set framework, but also paves the way for 
developing more advanced schemes by combining beyond-DFT approaches like hybrid functionals \cite{Levchenko2015,Lin2020,Lin2021} 
or $GW$ \cite{Ren2021}, recently available in NAO-based DFT codes, with the DMFT machinery.

\acknowledgements
\par This work is supported by National Natural Science Foundation of China (Grant Nos. 12134012, 11874335, 11874263),
the Strategic Priority Research Program of Chinese Academy of Sciences (Grant No. XDPB25), and The Max Planck
Parter Group for \textit{Advanced Electronic Structure Methods}.  
We thank valuable help from Li Huang on the impurity solver iQIST \cite{Huang2015,Huang2017,iQIST}, 
E. Gull and S. Iskakov on the impurity solver ALPS-CTHYB-SEGMENTS \cite{Gull2011b,Hafermann2013,ALPS-CTHYB-SEGMENTS},  
H. Shinaoka on the impurity solver ALPS-CTHYB \cite{Shinaoka2017b,ALPS-CTHYB}, and M. J. Han and J. H. Sim on 
the DFT+DMFT package DMFT-pack \cite{Sim2019, DMFTpack}.

\bibliographystyle{unsrt}
\bibliography{Reference_library}

\end{document}